\documentclass[onecolumn,aps,prd,preprintnumbers,showpacs,superscriptaddress,nofootinbib,amsmath,amssymb,floats,floatfix,showkeys,notitlepage,longbibliography]{revtex4-1}

\usepackage{comment}
\usepackage{graphicx}
\usepackage{subfigure}
\usepackage{palatino}
\usepackage[commandnameprefix=always]{changes}
\usepackage{hyperref}
\hypersetup{colorlinks=true,linkcolor=blue,urlcolor=blue,citecolor=blue}
\usepackage[toc,page]{appendix}
\usepackage[normalem]{ulem}

\usepackage{orcidlink}
\usepackage{lipsum}
\usepackage{graphicx}
\usepackage{subfigure}
\usepackage{palatino}
\usepackage{sans}
\usepackage{adjustbox}
\usepackage{latexsym}
\usepackage{amsmath}
\usepackage{amssymb}
\usepackage{amsfonts}
\usepackage{dcolumn}
\usepackage{bm}
\usepackage{tikz}
\usepackage{bigints}
\usepackage{array,tabularx,multirow,booktabs}
\usepackage[tracking=true]{microtype}
\SetTracking{}{500}
\SetTracking{encoding={*}, shape=sc}{40}
\UseRawInputEncoding 
\allowdisplaybreaks
\usepackage{adjustbox}
\usepackage{latexsym}
\usepackage{amsmath}
\usepackage{amssymb}
\usepackage{amsfonts}
\usepackage{dcolumn}
\usepackage{bm}
\usepackage{tikz}
\usepackage{bigints}
\usepackage{array,tabularx,multirow,booktabs}
\usepackage[tracking=true]{microtype}
\usepackage{color}
\UseRawInputEncoding 
\allowdisplaybreaks
\usepackage{soul} 

\def\0{{\sst{(0)}}}
\def\1{{\sst{(1)}}}
\def\2{{\sst{(2)}}}
\def\3{{\sst{(3)}}}
\def\4{{\sst{(4)}}}
\def\5{{\sst{(5)}}}
\def\6{{\sst{(6)}}}
\def\7{{\sst{(7)}}}
\def\8{{\sst{(8)}}}
\def\sst#1{{\scriptscriptstyle #1}}

\begin{document} \sloppy

\title{Asymptotically-Flat Black Hole Solutions in Symmergent Gravity \footnote{Dedicated to Durmu\c{s} Demir (1967-2024), our supervisor, candid friend, and guiding light. His unwavering supports always led our way.}}

\author{Beyhan Puli\c ce
}
\email{beyhan.pulice@sabanciuniv.edu}
\affiliation{Faculty of Engineering and Natural Sciences, Sabanc{\i} University, 34956 \.{I}stanbul, Turkey}

\author{Reggie C. Pantig
}
\email{rcpantig@mapua.edu.ph}
\affiliation{Physics Department, Map\'ua University, 658 Muralla St., Intramuros, Manila 1002, Philippines}

\author{Ali \"Ovg\"un
}
\email{ali.ovgun@emu.edu.tr}
\affiliation{Physics Department, Eastern Mediterranean
University, Famagusta, 99628 North Cyprus, via Mersin 10, Turkey}

\author{Durmu\c s~Demir
}
\email{durmus.demir@sabanciuniv.edu}
\affiliation{Faculty of Engineering and Natural Sciences, Sabanc{\i} University, 34956 \.{I}stanbul, Turkey}

\begin{abstract}
Symmergent gravity is an emergent gravity model with an $R+R^2$ curvature sector and an extended particle sector having new particles beyond the known ones. With constant scalar curvature, asymptotically flat black hole solutions are known to have no sensitivity to the quadratic curvature term (coefficient of $R^2$). With variable scalar curvature, however, asymptotically-flat symmergent black hole solutions turn out to explicitly depend on the quadratic curvature term. In the present work, we construct asymptotically-flat symmergent black holes with variable scalar curvature and use its evaporation, shadow, and deflection angle to constrain the symmergent gravity parameters. Concerning their evaporation, we find that the new particles predicted by symmergent gravity, even if they do not interact with the known particles, can enhance the black hole evaporation rate. Concerning their shadow, we show that statistically significant symmergent effects are reached at the $2\,\sigma$ level for the observational data of the Event Horizon Telescope (EHT) on the Sagittarius A* supermassive black hole. Concerning their weak deflection angle, we reveal discernible features for the boson-fermion number differences, particularly at large impact parameters. These findings hold the potential to serve as theoretical predictions for future observations and investigations on black hole properties.
\end{abstract}

\date{\today}



\maketitle

\section{Introduction}

In the Wilsonian sense, quantum field theories (QFTs) are characterized by a classical action and an ultraviolet (UV) cutoff $\Lambda$. Quantum loops lead to effective QFTs with loop momenta cut at $\Lambda$. Effective QFTs generally suffer from UV over-sensitivity problems: the scalar and gauge boson masses receive $\mathcal{O}(\Lambda^2)$ corrections. The vacuum energy, on the other hand, gets corrected by ${\mathcal{O}}(\Lambda^4)$  and ${\mathcal{O}}(\Lambda^2)$  terms. The gauge symmetries get explicitly broken. The question is simple: Can gravity emerge in a way that restores the explicitly broken gauge symmetries? Asking differently, can gravity emerge in a way alleviates the UV over-sensitivities of the effective QFT? This question has been answered affirmatively \cite{demir0,demir1}, with the constructions of a gauge symmetry-restoring emergent gravity model (see also \cite{demir2,demir3}). This model, briefly called as {\it symmergent gravity}, has been built by the observation that, in parallel with the introduction of Higgs field to restore gauge symmetry for a massive vector boson (with Casimir invariant mass) \cite{anderson,englert,higgs}, spacetime affine curvature can be introduced to restore gauge symmetries for gauge bosons with loop-induced (Casimir non-invariant) masses proportional to the UV cutoff $\Lambda$ \cite{demir0,demir1}. Symmergent gravity is emergent general relativity (GR) with a quadratic curvature term ($R+R^2$ gravity) and new particles beyond the known ones. Its curvature (as well as particle) sector exhibits distinctive signatures, as revealed in recent works on inflation \cite{irfan} and static black hole spacetimes \cite{Symmergent-bh,Symmergent-bh2,Symmergent-bh3,symmergentresults}.

Scientists predicted the existence of shadow of black holes a long time ago \cite{Synge:1966okc,Luminet:1979nyg,Falcke:1999pj}, but it wasn't until recently that the Event Horizon Telescope (EHT) team was able to acquire the first direct photograph of a black hole \cite{EventHorizonTelescope:2019pgp}. This photograph specifically depicted the shadow of M87*, a supermassive black hole at the center of the Messier 87 galaxy and then and Sagittarius A*  \cite{EventHorizonTelescope:2019pgp,EventHorizonTelescope:2022wkp}.  This remarkable discovery implies that we may soon be able to obtain valuable information on black holes, such as their mass, spin, and charge, in a systematic manner \cite{Ghosh:2020spb,Allahyari:2019jqz}. We can gather data to derive these characteristics by studying black hole shadows, which are essentially dark representations of their event horizon, and photon rings, which are dazzling images formed by photons circling around black holes \cite{Bambi:2019tjh,Vagnozzi:2022moj,Prashant2021}, which are especially noteworthy since they provide accurate gravity constraints in the strong-field regime \cite{Ovgun:2018tua,Ovgun:2020gjz,Ovgun:2019jdo,Kuang:2022xjp,Kumaran:2022soh,Mustafa:2022xod,Okyay:2021nnh,Atamurotov:2022knb,Abdikamalov:2019ztb,Abdujabbarov:2016efm,Atamurotov:2015nra,Papnoi:2014aaa,Abdujabbarov:2012bn,Atamurotov:2013sca,Cunha:2018acu,Gralla:2019xty,Belhaj:2020okh,Belhaj:2020rdb,Konoplya2019,Wei2019,Ling:2021vgk,Kumar:2020hgm,Kumar2017EPJC,Cunha:2016wzk,Cunha:2016bpi,Cunha:2016bjh,Zakharov:2014lqa,Tsukamoto:2017fxq,Chakhchi:2022fls,Li2020,EventHorizonTelescope:2021dqv,Vagnozzi:2022moj}. These discrepancies could be caused by many reasons linked with several alternative theories of gravity \cite{Pantig:2022ely,Pantig:2022gih,Lobos:2022jsz,Uniyal:2022vdu,Ovgun:2023ego,Uniyal:2023inx,Panotopoulos:2021tkk,Panotopoulos:2022bky,Khodadi:2022pqh,Khodadi:2021gbc,Meng:2023unt,Shaikh:2022ivr,Shaikh:2021cvl,Shaikh:2018lcc,Shaikh:2019hbm} or the surrounding astrophysical conditions in which the black hole is located \cite{Pantig:2022whj,Pantig:2022sjb, Pantig:2023yer, Wang:2019skw, Roy:2020dyy,Konoplya:2021ube,Anjum:2023axh,Hou:2018bar}. Therefore, it becomes crucial to investigate modified gravity theories and establish constraints by utilizing the black hole's shadow alongside astrophysical data, such as observations from telescopes like EHT.

In the present work, we construct and analyze asymptotically-flat, static, spherically-symmetric symmergent gravity black holes. In the case of constant scalar curvature, the quadratic curvature term (coefficient of $R^2$) is known not to affect the asymptotically flat spacetimes \cite{constant-R-1,constant-R-2,constant-R-3}. In the case of variable scalar curvature, however, there arise asymptotically-flat solutions with explicit dependence on the quadratic curvature term  \cite{nguyen3} (see also \cite{nguyen1,nguyen2}). In Sec. II, we give a detailed discussion of the symmergent gravity \cite{demir0,demir1} in regarding its new particles sector (Sec. IIA) and its curvature sector (Sec. IIB). In Sec. III, we construct asymptotically-flat, static, spherically symmergent gravity black holes with variable scalar curvature. Our analysis goes beyond \cite{nguyen3} as we consider both positive and negative values of the boson-fermion number difference (or the parameter $\gamma$ in \cite{nguyen3}). In both cases, we demonstrate asymptotic flatness of the metric, with approximate analytic calculations and the exact numerical solutions. In Sec. IV, we compute the Hawking temperature using the tunneling method and state that black hole evaporation could be accelerated if there exist considerable new light particles. In Sec. V, we analyze how the symmergent black hole can be probed via its shadow cast and weak deflection angle. And we have shown that the symmergent effects on the shadow are small, and the weak deflection angle can distinguish different boson-fermion number differences at fairly large impact factors. In Sec. VI, we conclude. 

\section{Symmergent Gravity} \label{sec2}

In this section, we give a brief description of symmergent gravity in terms of its fundamental parameters. The starting point is quantum field theories (QFTs). Quantum fields are endowed with mass and spin as their Casimir invariants of the Poincar\'e group. Fundamentally, QFTs are intrinsic to the flat spacetime simply because they rest on a Poincar\'e-invariant (translation-invariant) vacuum state \cite{incompatible,wald}.  Flat spacetime means the total absence of gravity, and its incorporation necessitates the QFTs to be carried into curved spacetime. But this carriage is hampered by Poincar\'e breaking in curved spacetime \cite{dyson,wald}, and this hamper and the absence of a quantum theory of gravity \cite{thooft}  together lead one to emergent gravity framework \cite{sakharov,visser,verlinde} as a viable approach.

In general, the loss of Poincar\'e invariance could be interpreted as the emergence of gravity form within the QFT \cite{fn2}. In a QFT, curvature can emerge at the Poincar\'e breaking sources. One natural Poincar\'e breaking source is the hard momentum cutoff on the QFT. Indeed, an ultraviolet (UV)  cutoff $\Lambda$ \cite{cutoff} limits momenta $p_\mu$ within $-\Lambda^2 \leq \eta^{\mu\nu} p_\mu p_\nu \leq \Lambda^2$ interval as the intrinsic validity edge of the QFT \cite{cutoff}. Under the loop corrections under the cutoff $\Lambda$, the  action $S[\eta,\phi,V]$ of a QFT of scalars $S$ and gauge bosons $V_\mu$ receives the correction (with $(+,-,-,-)$ metric signature appropriate for QFTs)
\begin{eqnarray}
 \delta S[\eta,\phi,V]  = \int d^4 x \sqrt{-\eta} \left\{-c_{\rm O} \Lambda^4 - \sum_i c_{m_i} m_i^2 \Lambda^2 - c_S \Lambda^2 S^\dagger S + c_V \Lambda^2 V_{\mu} V^{\mu} \right\},
 \label{action-0}
\end{eqnarray}
in which $\eta_{\mu\nu}$ is the flat metric, $m_i$ stands for the mass of a QFT field $\psi_i$ (summing over all the fermions and bosons),  and $c_{\rm O}$, $c_m$, $c_S$ and $c_V$ are respectively the loop factors describing the quartic vacuum energy correction, quadratic vacuum energy correction, quadratic scalar mass correction, and the loop-induced gauge boson mass \cite{gauge-break1,gauge-break2}. As revealed by the gauge boson mass term $c_V \Lambda^2 V_{\mu} V^{\mu}$, the UV cutoff $\Lambda$ breaks gauge symmetries explicitly since $\Lambda$ is not a particle mass, that is, $\Lambda$ is not a Casimir invariant of the Poincar\'e group. The loop factor $c_V$ (and $c_S$) depends on the details of the QFT. (It has been calculated for the standard model gauge group in \cite{demir1,demir2}.)

In Sakharov's induced gravity \cite{sakharov,visser}, the UV cutoff $\Lambda$ is associated with the Planck scale, albeit with explicitly broken gauge symmetries and Planckian-size cosmological constant and scalar masses. In recent years, Sakharov's setup has been approached from a new perspective in which priority is given to the prevention of the explicit gauge symmetry breaking \cite{demir0,demir1}. For this aim, one first takes the effective QFT in (\ref{action-0}) to curved spacetime of a metric $g_{\mu\nu}$ such that the gauge boson mass term is mapped as $c_V \Lambda^2 \eta^{\mu\nu} V_{\mu} V_{\nu} \longrightarrow c_V V_{\mu} \left(\Lambda^2 g^{\mu\nu}-R^{\mu\nu}(g)\right)V_\nu$ in agreement with the fact that the Ricci curvature $R^{\mu\nu}(g)$ of the metric $g_{\mu\nu}$ can arise only in the gauge sector via the covariant derivatives \cite{demir0,demir1,demir2}. One next inspires from the Higgs mechanism to promote the UV cutoff $\Lambda$ to an appropriate spurion field. Indeed, in parallel with the introduction of the Higgs field to restore gauge symmetry for a massive vector boson (Poincare-conserving mass) \cite{anderson,englert,higgs}, one can introduce spacetime affine curvature to restore gauge symmetries for gauge bosons with loop-induced (Poincare-breaking) masses proportional to $\Lambda$ \cite{affine1,affine2,demir0,demir1}. Then, one is led to the map \begin{eqnarray}
\Lambda^2 g^{\mu\nu} \rightarrow {\mathbb{R}}^{\mu\nu}(\Gamma),
\label{map}
\end{eqnarray}
in which ${\mathbb{R}}^{\mu\nu}(\Gamma)$ is the Ricci curvature of the affine connection $\Gamma^{\lambda}_{\mu\nu}$, which is completely independent of the curved metric $g_{\mu\nu}$ and its Levi-Civita connection \cite{affine1,affine2,affine3}. This map is analog of the map $M_V^2 \rightarrow \phi^\dagger \phi$ of the vector boson mass $M_V$ (Poincare conserving) into the Higgs field $\phi$. {\color{black} Under the map (\ref{map}), the effective QFT in (\ref{action-0}) takes the form
\begin{eqnarray}
 \delta S[g,\phi,V, {\mathbb{R}}]  = \int d^4 x \sqrt{-\eta} \left\{-\frac{c_{\rm O}}{16}  {\mathbb{R}}^2 - \sum_i \frac{c_{m_i}}{4} m_i^2 {\mathbb{R}} - \frac{c_S}{4} {\mathbb{R}} S^\dagger S + c_V V_{\mu} \left({\mathbb{R}}^{\mu\nu}(\Gamma)-R^{\mu\nu}(g)\right)V_\nu \right\},
 \label{action-1}
\end{eqnarray}
in which ${\mathbb{R}} \equiv g^{\mu\nu} {\mathbb{R}}_{\mu\nu}(\Gamma)$ is the affine scalar curvature \cite{demir1}, which tends to the usual metrical scalar curvature under the affine dynamics to be discussed in Sec.\ref{subsec: new gravity} below}. This metric-Palatini theory contains both the metrical curvature $R(g)$ and the affine curvature ${\mathbb{R}}(\Gamma)$. From the second term, Newton's gravitational constant $G_N$ is read out to be 
\begin{eqnarray}
G_N^{-1}= 4 \pi \sum_i c_{m_i} m_i^2 \xrightarrow{\rm one\ loop} \frac{1}{8\pi} {\rm str}\!\left[{\mathcal{M}}^2 \right],
\label{MPl}
\end{eqnarray}
where ${\mathcal{M}}^2$ is the mass-squared matrix of all the fields in the QFT spectrum. In the one-loop expression, ${\rm str}[\dots]$ stands for super-trace namely ${\rm str}[{\mathcal{M}}^2] = \sum_i (-1)^{2s_i} (2 s_i +1) {\rm tr}[{\mathcal{M}}^2]_{s_i}$ in which ${\rm tr}[\dots]$
is the usual trace (including the color degrees of freedom), $s_i$ is the spin of the QFT field $\psi_i$  ($s_i=0,1/2,\dots$), and $[{\mathcal{M}}^2]_{s_i}$ is the mass-squared matrix of the fields having that spin (like mass-squared matrices of scalars ($s_i=0$), fermions $(s_i=1/2)$ and so on). One keeps in mind that ${\rm tr}[\dots]$ encodes degrees of freedom $g_i$ (like color and other degrees of freedom) of the particles.

\subsection{First Prediction of Symmergence: Naturally-Coupled New Particles}
\label{subsec: new particles}
It is clear that the known particles (the quarks, leptons, gauge bosons, and the Higgs boson in the Standard Model (SM)) can generate Newton's constant in (\ref{MPl}) 
neither in size (${\rm str}\!\left[{\mathcal{M}}^2 \right] \sim ({\rm TeV})^2$ in the SM) nor in sign (${\rm str}\!\left[{\mathcal{M}}^2 \right] < 0$ in the SM). It is thus necessary to introduce new particles. What is interesting about these new particles is that they do not have to couple to the known SM particles since the only constraint on them is the super-trace in (\ref{MPl}) \cite{demir0,demir1}. They may not interact with the SM particles, but they may interact, too. If they interact, there is no symmetry principle or selection rule that forces their coupling strengths to be of the SM size (as in supersymmetry, extra dimensions, and technicolor). To see this, it suffices to consider renormalizable interactions of the form 
\begin{eqnarray}
{\mathcal{L}}_{int} = \lambda_{H H^\prime} \left(H^\dagger H\right)\cdot \left(H^{\prime\dagger}H^\prime\right),
\label{int-0}
\end{eqnarray}
in which $H$ is the SM Higgs field, and $H^\prime$ stands for new scalar fields. The main point is that the coupling constant $\lambda_{H H^\prime}$ is forced to take a value $\lambda_{H H^\prime}\sim \lambda_{SM}$ in supersymmetry, extra dimensions, and technicolor due to their symmetry structures correlating the SM fields and the new fields. (In supersymmetry, for instance, superpartners are the new fields that interact with the SM fields with SM-sized couplings due to supersymmetry transformations.)   Here, $\lambda_{SM}$ is a typical SM coupling such as the top Yukawa coupling $\lambda_t \approx 1$, and under such sizable couplings, the $H^\prime$-loop corrections to the Higgs mass  
\begin{eqnarray}
 \delta m_H^2 \propto \lambda_{H H^\prime} m^2_{H^\prime} \log \frac{m^2_{H^\prime}} {m^2_{H}},     
 \label{corr-0}
\end{eqnarray}
become too large to remain within the experimental error bars when $m_{H^\prime}\gg m_{H}$. As a matter of fact, supersymmetry and other new physics models have been sidelined based on these corrections in that the LHC experiments seem to be near the $m_{H^\prime}\gg m_{H}$ regime run after run. 

Symmergence is entirely different. The reason is that there is no symmetry structure correlating the SM fields and the new fields, so the coupling constant  $\lambda_{H H^\prime}$ in (\ref{int-0}) can take any perturbative value: $\lambda_{H H^\prime}=0\cdots \lambda_{SM}$. These new fields we call {\it symmerons} to distinguish them from new physics models with sizable couplings to the SM \cite{demir0,demir1}. Broadly speaking, symmerons can come in three classes \cite{dark-sector1,demir0,demir1,demir2,demir3}:
\begin{enumerate}
    \item {\it Visible Symmerons} are those new particles that are endowed with the SM charges and are allowed, therefore to interact with the SM with sizable couplings (like $\lambda_{H H^\prime}\sim \lambda_{SM}$). These particles weigh necessarily at the SM scale if they are not to destabilize the SM Higgs sector via the Higgs mass corrections they induce  (like $\delta m_H^2$). The LHC results and results from various dark matter searches seem to disfavor such symmerons. 

    \item {\it Dark Symmerons} are those new particles that are neutral under the SM charges and are allowed therefore to interact with the SM with weak (or feeble) couplings in the range $\lambda_{H H^\prime} \lesssim \lambda_{max}$ within which the $H^\prime$-loop corrections in (\ref{corr-0}) remain small enough to keep the SM Higgs sector stable. In fact, as suggested by (\ref{corr-0}), one can take   $\lambda_{max} \simeq \frac{m_H^2}{m^2_{H^\prime}}$ as an optimal coupling strength for such symmerons. 

    \item {\it Black Symmerons} are those new particles that do not couple to the SM at all (like $\lambda_{H H^\prime}=0$). These symmerons form a secluded sector that interacts with the SM via only gravity. This kind of symmerons (and the feebly-coupled dark symmerons) broadly agree with most observations.  
\end{enumerate}
The built-in immunity of symmergence to coupling strengths between the SM and the requisite new fields is an important property. This picture is in overall agreement with the existing astrophysical \cite{dark-sector2,dark-sector3}, cosmological \cite{dark-sector-cosmo} and collider \cite{dark-sector-LHC} searches (see also \cite{dark-sector1}). The symmerons are a prediction of symmergence, and we shall discuss their impact on black hole physics in the next section. 

\subsection{Second Prediction of Symmergence: $R+R^2$ Gravity with Loop-Induced Parameters}
\label{subsec: new gravity}

The action (\ref{action-1}) remains stationary against variations in the affine connection provided that 
\begin{eqnarray}
\label{gamma-eom}
{}^{\Gamma}\nabla_{\lambda} {\mathbb{D}}_{\mu\nu} = 0,
\end{eqnarray}
such that ${}^{\Gamma}\nabla_{\lambda}$ is the covariant derivative of the affine connection $\Gamma^\lambda_{\mu\nu}$, and 
\begin{eqnarray}
\label{q-tensor}
{\mathbb{D}}_{\mu\nu} = \left(\frac{1}{16\pi G} +   \frac{c_S}{4} S^\dagger S + \frac{c_{\rm O}}{8} g^{\alpha\beta} {\mathbb{R}}_{\alpha\beta}(\Gamma)\right) g_{\mu\nu} - c_{V} V_{\mu}V_{\nu},
\end{eqnarray}
is the field-dependent metric.  The motion equation (\ref{gamma-eom}) implies that ${\mathbb{D}}_{\mu\nu}$ is covariantly-constant with respect to $\Gamma^\lambda_{\mu\nu}$. In solving (\ref{gamma-eom}),  it is legitimate to make the expansions 
\begin{eqnarray}
\Gamma^{\lambda}_{\mu\nu}&=&{}^{g}\Gamma^{\lambda}_{\mu\nu} + {\mathcal{O}}\left(G_N\right),
\label{expand-conn}
\end{eqnarray}
and
\begin{eqnarray}
{\mathbb{R}}_{\mu\nu}(\Gamma) &=& R_{\mu\nu}(g) +{\mathcal{O}}\left(G_N\right),
\label{expand-curv}
\end{eqnarray}
since the Planck scale in (\ref{MPl}) is the largest scale. In these expansions, though not made explicit, both $\Gamma^{\lambda}_{\mu\nu}$ and ${\mathbb{R}}_{\mu\nu}(\Gamma)$ contain  pure derivative terms  at  the next-to-leading ${\mathcal{O}}\left(G_N\right)$ order \cite{demir2,demir3}. The expansion in (\ref{expand-conn}) ensures that the affine connection $\Gamma^{\lambda}_{\mu\nu}$ is solved algebraically order by order in $G_N$  despite the fact that its motion equation (\ref{gamma-eom}) involves its own curvature ${\mathbb{R}}_{\mu\nu}(\Gamma)$ through ${\mathbb{D}}_{\mu\nu}$   \cite{affine1,affine2}. The expansion (\ref{expand-curv}), on the other hand, ensures that the affine curvature  ${\mathbb{R}}_{\mu\nu}(\Gamma)$ is equal to the metrical curvature $R_{\mu\nu}(g)$ up to a doubly-Planck suppressed remainder. In essence, what happened is that the affine dynamics took the affine curvature ${\mathbb{R}}$ from its UV value $\Lambda_\wp^2$  in (\ref{map}) to its IR value $R$ in (\ref{expand-curv}). This way, the GR emerges holographically \cite{holog1,holog2} via the affine dynamics such that loop-induced gauge boson masses get erased, and scalar masses get stabilized by the curvature terms. This mechanism renders effective field theories natural regarding their destabilizing UV sensitivities \cite{demir0,demir1}. It gives rise to a new framework in which {\it (i)} the gravity sector is composed of the Einstein-Hilbert term plus a curvature-squared term, and {\it (ii)} the matter sector is described by an ${\overline{MS}}$-renormalized QFT \cite{demir0,demir1,demir2}. We call this framework gauge symmetry-restoring emergent gravity or simply {\it symmergent gravity} to distinguish it from other emergent or induced gravity theories in the literature. 

It is worth noting that symmergent gravity is not a loop-induced curvature sector in curved spacetime  \cite{visser,birrel}. In contrast, symmergent gravity arises when the flat spacetime effective QFT is taken to curved spacetime \cite{demir0,demir1,demir2} in a way restoring the gauge symmetries broken explicitly by the UV cutoff. All of its couplings are loop-induced parameters deriving from the particle spectrum of the QFT (numbers and masses of particles). It is with these loop features that the GR 
emerges. In fact, the metric-Palatini action (\ref{action-1}) reduces to the metrical gravity theory 
\begin{eqnarray}
     \int d^4x \sqrt{-g} \Bigg\{\! &&
 -\frac{{\mathbb{R}}(g)}{16\pi G_N} - \frac{c_{\rm O}}{16} \left({\mathbb{R}}(g)\right)^2 +\frac{c_S}{4}  S^\dagger S\, {\mathbb{R}}(g) + c_V {\rm tr}\left[V^{\mu}\!\left({\mathbb{R}}_{\mu\nu}(\Gamma)- R_{\mu\nu}({}^g\Gamma)\right)\!V^{\nu}\right]\! \Bigg\}\nonumber\\
&&\xrightarrow{\rm equation\, (\ref{expand-curv})}
\int d^4x \sqrt{-g} \Bigg\{\!
-\frac{R}{16\pi G_N}   - \frac{c_{\rm O}}{16} R^2  -\frac{c_S}{4} S^\dagger S  R + {\mathcal{O}}\!\left(G_N\right)\!\Bigg\},
\label{reduce-nongauge}
\end{eqnarray}
after replacing the affine curvature ${\mathbb{R}}_{\mu\nu}(\Gamma)$ with its solution in (\ref{expand-curv}). Of the parameters of this emergent GR action, Newton's constant $G_N$ was already defined in (\ref{MPl}). The loop factor $c_S$ depends on the underlying QFT. (It reads $c_S\simeq 0.29$ in the standard model.) The loop factor $c_{\rm O}$, which was associated with the quartic ($\Lambda^4$) corrections in the flat spacetime effective QFT in (\ref{action-0}), turned to the coefficient of quadratic-curvature $(R^2)$ term in the symmergent GR action in (\ref{reduce-nongauge}). At one loop, it takes the value
\begin{eqnarray}
c_{\rm O} = \frac{n_B - n_F}{128 \pi^2},
\label{param1}
\end{eqnarray}
in which  $n_B$ ($n_F$) stands for the total number of bosonic (fermionic) degrees of freedom in the underlying QFT (including the color degrees of freedom). Both the $n_B$ bosons and $n_F$ fermions contain not only the known standard model particles but also the completely new particles. As was commented just above (\ref{gamma-eom}), it is a virtue of symmergence that these new particles do not have to couple to the known ones non-gravitationally. 

In the above, we have consistently dropped the total vacuum energy, assuming that the cosmological constant problem has been solved by an appropriate method. Saying differently, we have assumed that the tree-level vacuum energy and logarithmic loop corrections left over (power-law $\Lambda^4$ corrections have been converted to curvature) from symmergence have been alleviated by an appropriate mechanism. 

Symmergence makes gravity emerge from within the flat spacetime effective QFT. Fundamentally, as follows from the action (\ref{action-0}), Newton's constant $G_N$ in (\ref{MPl}) and the quadratic curvature coefficient $c_{\rm O}$ in (\ref{param1}) transpired in the flat spacetime effective QFT from the matter loops. (The loop factor $c_S$ and similar parameters involve matter fields.)  A  glance at the second line of (\ref{reduce-nongauge}) reveals that symmergent gravity is an $R+R^2$ gravity theory with the non-zero cosmological constant. In fact, it can be put in the form  
\begin{eqnarray}
S=-\frac{c_{\rm O}}{16} \int d^4 x \sqrt{-g}\left(R^2 + 6 \gamma G_N^{-1} R\right),
\label{fr-action}
\end{eqnarray}
after leaving aside the scalars $S$ and the other matter fields, after switching to $(-,+,+,+)$ metric signature (appropriate for the black hole analysis in the sequel), and after introducing the constant
\begin{eqnarray}
\gamma = - \frac{1}{6\pi c_ {\rm O}}=-\frac{64 \pi}{3 (n_B - n_F)},
\label{gamma}
\end{eqnarray}
where $c_{\rm O}$ was defined in equation (\ref{param1}) above. Here, one recalls that $(n_B-n_F)_{SM}=-62$ and hence $(\gamma)_{SM}\approx 1$. However, as already discussed in Sec. \ref{subsec: new particles},  
the SM spectrum is insufficient for inducing the gravitational constant in (\ref{MPl}), and thus, it is necessary to introduce new fields beyond the SM. The new fields cause {\it (i)} Newton's constant to be induced as in (\ref{MPl}), {\it (ii)} $|n_B-n_F|$ to be larger or smaller than $|(n_B-n_F)_{SM}|$ depending on the mass spectrum ($n_B-n_F=0$ corresponds to Bose-Fermi balance as in supersymmetric QFTs and nullifies the quadratic curvature term), and  {\it (ii)} ${\rm sign}(n_B-n_F)$ to be positive ($n_B>n_F$) or negative ($n_B<n_F$) depending again on the mass spectrum. All this means that $\gamma$ can deviate from its SM value significantly in both the positive and negative directions. 

\section{Asymptotically-Flat Black Hole Spacetime in Symmergent Gravity}
The symmergent gravitational action in (\ref{fr-action}) remains stationary against variations in the metric provided that the Einstein field equations
\begin{eqnarray}
   (R + 3\gamma G_N^{-1})R_{\mu\nu}-\frac{1}{4} (R + 6 \gamma G_N^{-1}) R g_{\mu\nu} -\left(\nabla_\mu\nabla_\nu - \Box g_{\mu\nu}\right)R  = 0,
   \label{eins-eqn}
\end{eqnarray}
hold as motion equation for the metric $g_{\mu\nu}$. These equations possess static, spherically-symmetric, asymptotically-flat solutions. Such solutions with constant scalar curvature ($R=$ constant) turn out to be insensitive to the quadratic-curvature term in (\ref{fr-action}). In other words, such solutions bear no sensitivity to the loop factor $c_{\rm O}$, and they cannot thus probe vacuum symmergent gravity. (This insensitivity is a common feature of all $R+R^2$ gravity theories \cite{constant-R-1,constant-R-2,constant-R-3}, and gets  lost in the presence of the cosmological constant \cite{Symmergent-bh,Symmergent-bh2,Symmergent-bh3,symmergentresults}.)

The Einstein field equations (\ref{eins-eqn}) possess also static, spherically-symmetric, asymptotically-flat solutions with variable scalar curvature ($R=$ variable) \cite{nguyen3} (see also \cite{nguyen1,nguyen2}). These solutions show direct sensitivity to the quadratic-curvature term in (\ref{fr-action}). They are thus sensitive to $c_{\rm O}$, and have the ability to probe symmergent gravity. In fact, introducing the dimensionless black hole mass $M\rightarrow G_N^{1/2} M \equiv M$ and dimensionless radial distance $r\rightarrow G_N^{-1/2} r \equiv r$, $r\rightarrow G_N^{-1/2} r \equiv r$, {\color{black} recently it has been shown that the metric \cite{nguyen1}
\begin{eqnarray}
    ds^2 = -(1-\varphi(r))\Psi(r) dt^2 + \frac{d r^2}{(1+\varphi(r))\Psi(r)} + r^2 d\Omega^2, 
    \label{metric0}
\end{eqnarray}
satisfies the Einstein field equations (\ref{eins-eqn}) at the linear order in $\varphi(r)$ provided that the differential equation \cite{nguyen1} (see also \cite{nguyen2,nguyen3})
\begin{eqnarray}
    \left(r^2 \Psi(r) \varphi^\prime(r)\right)^\prime= \gamma r^2 \varphi(r),
    \label{varphi-eq}
\end{eqnarray}
is satisfied for both $\gamma>0$ (as was  assumed in \cite{nguyen1}) and $\gamma <0$ (as will be considered additionally in the present work). Here, the Schwarzschild lapse function 
\begin{eqnarray}
   \Psi(r) = 1 - \frac{2M}{r}, 
\end{eqnarray}
has a zero (horizon) at $r=2M$, as expected. Singularity structure of the spacetime (\ref{metric0}) is revealed by the Kretschmann scalar 
\begin{align}
R_{\mu\nu\sigma\rho}R^{\mu\nu\sigma\rho} = \frac{48 M^2}{r^6} 
\left(1
+2\varphi(r) \right),
\end{align}
which is seen to differ from the usual Schwarzschild value by $\varphi(r)$. }
\begin{figure*}[h!]
   \centering
    \includegraphics[width=0.48\textwidth]{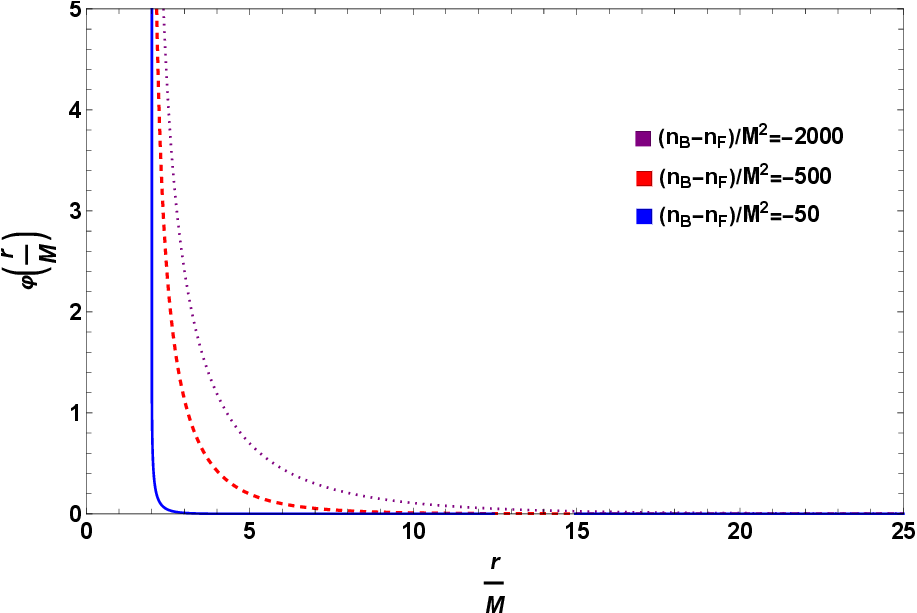} 
    \includegraphics[width=0.48\textwidth]{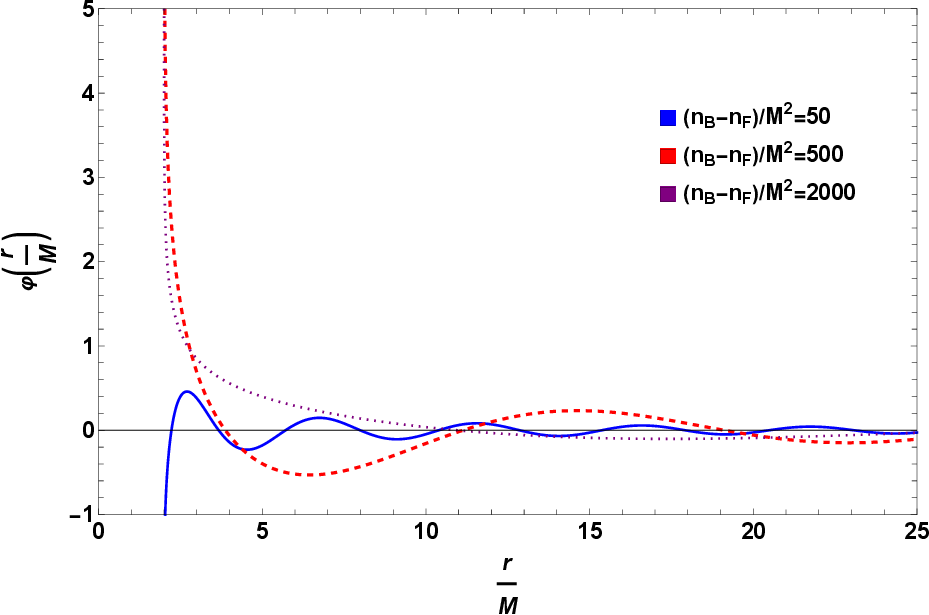}
    \caption{Exact numerical solution of $\varphi$ from the defining equation (\ref{varphi-eq}) as a function of the radial distance $r/M$ for selected negative values of $n_B - n_F$ ($\gamma >0$)(left panel) and positive values of $n_B - n_F$ ($\gamma < 0$)(right panel). {\color{black} In view of the validity domain of the metric (\ref{metric0}),  it is clear that essentially only $\varphi(r)$ at radial distances  $r\gtrsim 2$ are allowed (above the Schwarzschild horizon).}}
    \label{varphi_sols}
\end{figure*}

\begin{figure*}[h!]
   \centering
    \includegraphics[width=0.48\textwidth]{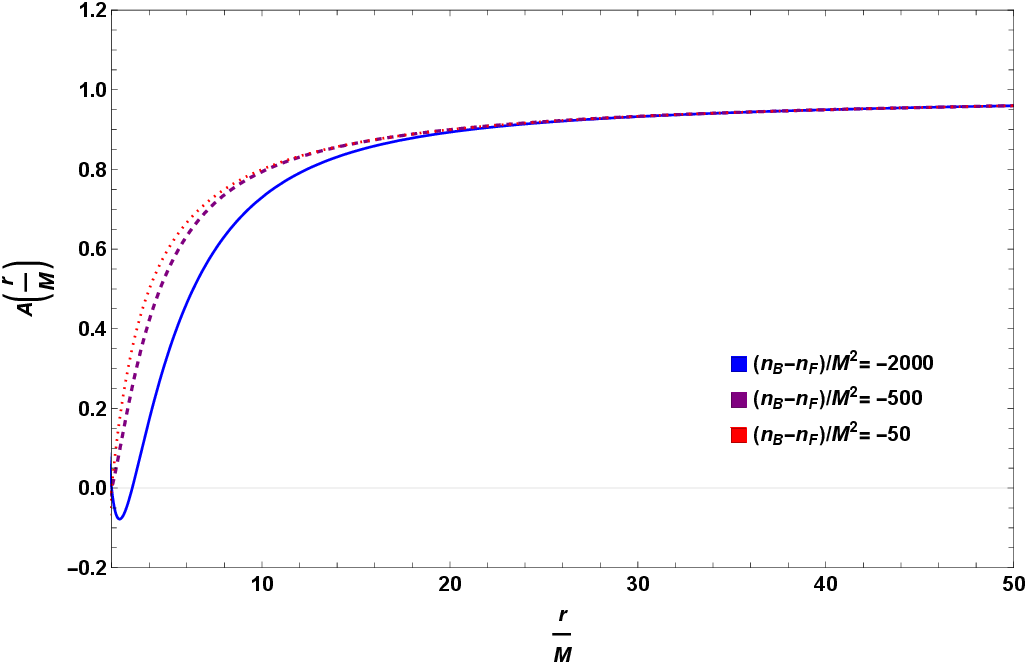} 
    \includegraphics[width=0.48\textwidth]{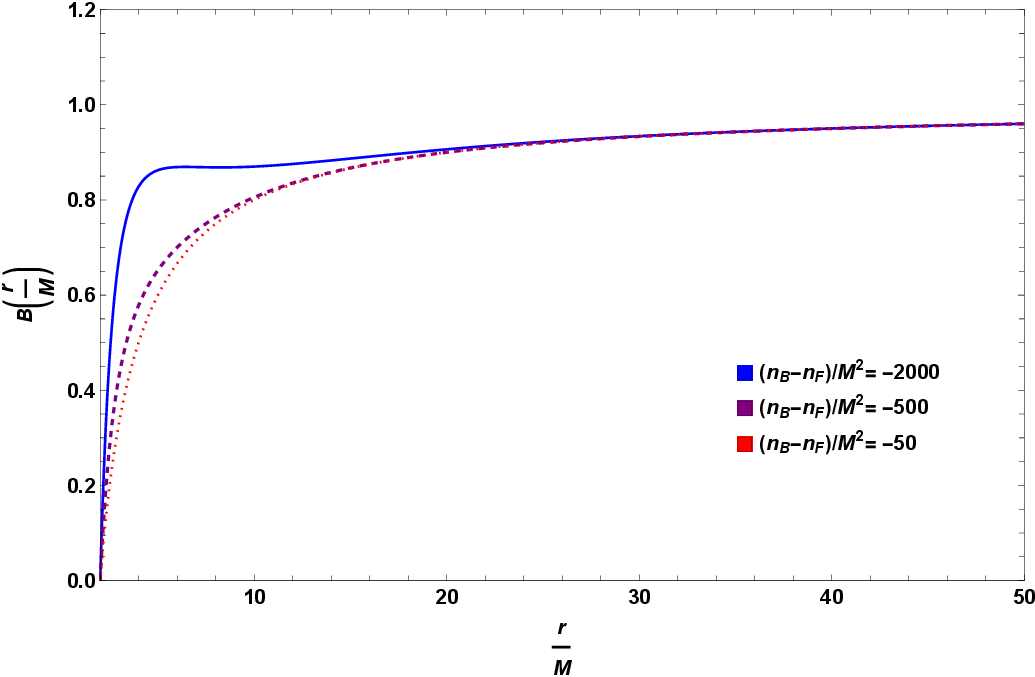}
    \caption{Exact solutions of the metric potentials (\ref{metric-A1}) and (\ref{metric-B1}) as functions of the radial distance $r$ for selected negative values of $(n_B - n_F)/M^2$ ($\gamma > 0$). {\color{black} These solutions are reliable mainly at distances $r\gtrsim 2$ in view of the validity of the metric (\ref{metric0}).} }
    \label{plot-A1B1}
\end{figure*}

\begin{figure}[h!]
   \centering
    \includegraphics[width=0.48\textwidth]{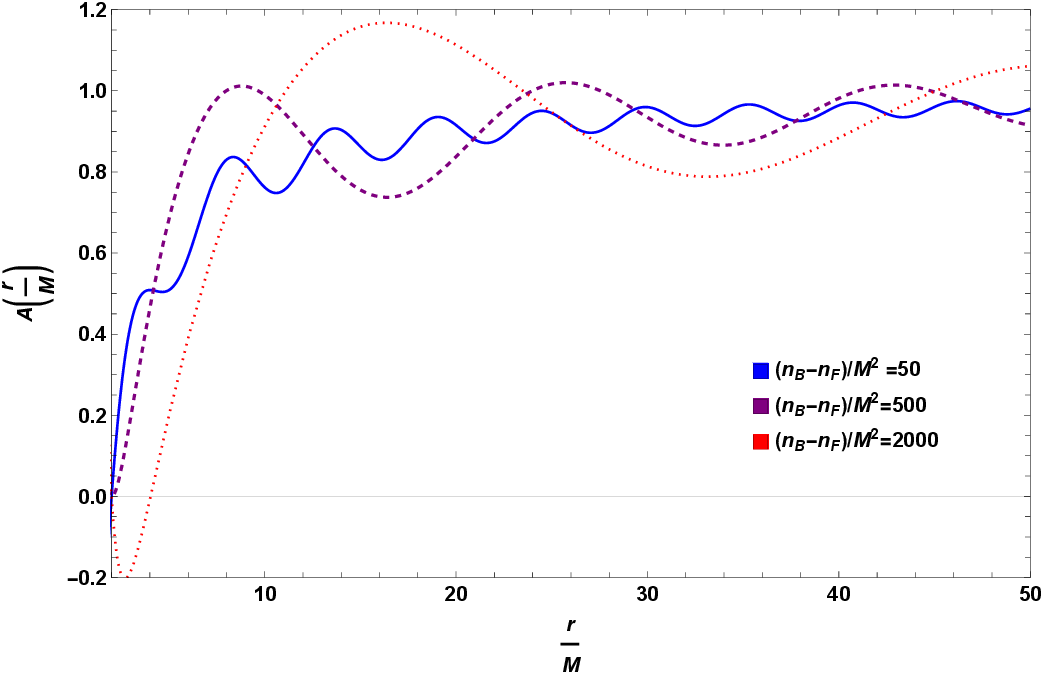} 
    \includegraphics[width=0.48\textwidth]{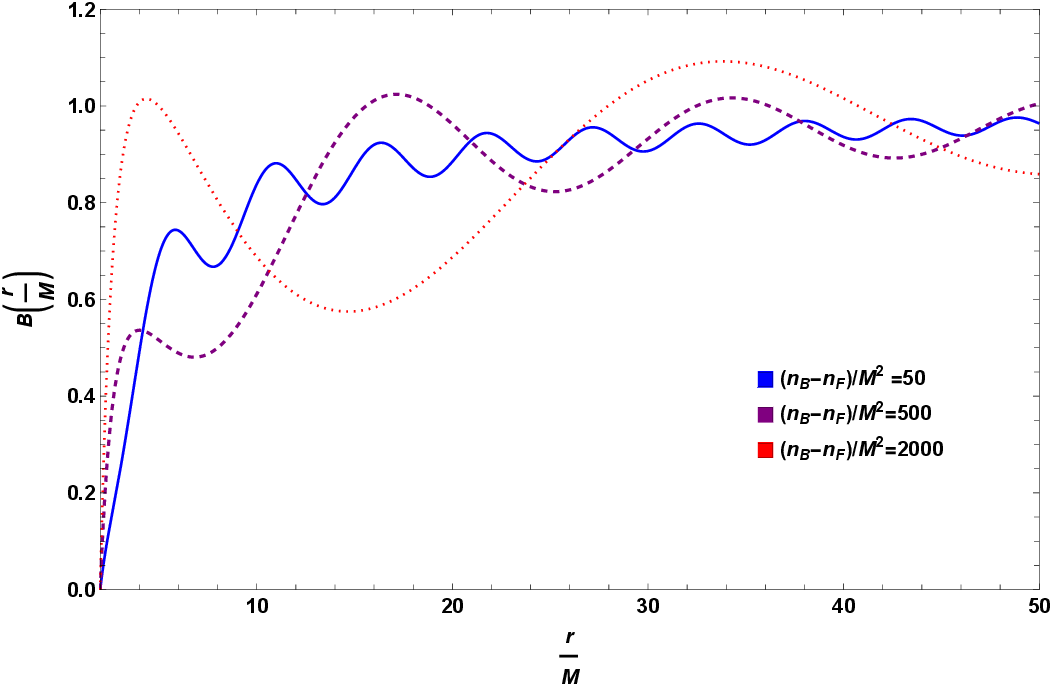}
    \caption{Exact solutions of the metric potentials (\ref{metric-A2}) and (\ref{metric-B2}) as functions of the radial distance $r$ for selected positive values of $(n_B - n_F)/M^2$ ($\gamma \leq 0$). {\color{black} These solutions are reliable mainly at distances $r\gtrsim 2$ in view of the validity of the metric (\ref{metric0}).}}
    \label{plot-A2B2}
\end{figure}

Our analyses below of various physical quantities rest on the exact solution of the $\varphi$-equation in (\ref{varphi-eq}) in which the Schwarzschild lapse function $\Psi(r)$ is treated exactly. In fact, the $\varphi(r)$ in Fig. \ref{varphi_sols} is obtained by such an exact numerical solution. {\color{black} For clarity and definiteness, however, it proves useful to illustrate these exact numerical solutions with approximate large-$r$ asymptotes. In this regard, large-$r$ behavior of $\varphi(r)$ is given by}
\begin{eqnarray}
    \varphi(r) = \begin{cases} \frac{e^{-\sqrt{\gamma} r}}{\sqrt{\gamma} r} & \gamma > 0\\ \\
     \frac{\cos{(\sqrt{|\gamma|} r)}}{\sqrt{|\gamma|}r} & \gamma < 0
    \end{cases}
    \label{phi-soln}
\end{eqnarray}
in the limit $r\gg 2M$ so that $\Psi(r)\approx 1$ in equation (\ref{varphi-eq}) above. The first line ($\gamma > 0$) of this equation was already derived in \cite{nguyen3}. The second line, our derivation, is also important since the parameter $\gamma$, as defined in  (\ref{gamma}), is positive for $n_B < n_F$
and negative for $n_B > n_F$. As revealed by (\ref{phi-soln}), the metric in (\ref{metric0}) is asymptotically-flat for both $\gamma>0$ and $\gamma\leq 0$ but its approach to the flatness is different in the two cases. In view of these asymptotic solutions, {\color{black} the metric (\ref{metric0}) can be put in a more definitive form as}
\begin{equation}
d s^{2}=-A(r) d t^{2}+ \frac{ d r^{2}}{B(r)}+C(r) \left ( d \theta^{2}+\sin^2\theta d \phi^{2} \right ),
\label{metric-antz}
\end{equation}
with the metric potentials
\begin{eqnarray}
\label{metric-A1}
A(r) &=& \left ( 1 -\frac{2M}{r}\right ) \left (1 - \frac{e^{- {r} \mathrm{\sqrt{\gamma}}}}{r \mathrm{\sqrt{\gamma}}} \right ), \\
\label{metric-B1}
B(r) &=& \left ( 1 -\frac{2M}{r}\right ) \left (1 + \frac{e^{- {r} \mathrm{\sqrt{\gamma}}}}{r \mathrm{\sqrt{\gamma}}} \right ), \\
\label{metric-C1}
C(r) &=& r^2 \left ( 1 - \frac{e^{- r\mathrm{\sqrt{\gamma}}}}{r \mathrm{\sqrt{\gamma}}} \right ).
\end{eqnarray}
for $\gamma > 0$, and 
\begin{eqnarray}
\label{metric-A2}
A(r) &=& \left ( 1-\frac{2 M}{r} \right ) \left ( 1 -\frac{\cos (\sqrt{|\gamma|} r)}{r \sqrt{|\gamma|}} \right ),\\
\label{metric-B2}
B(r) &=& \left ( 1-\frac{2 M}{r} \right ) \left ( 1 +\frac{\cos (\sqrt{|\gamma|} r)}{r \sqrt{|\gamma|}} \right ),\\
\label{metric-C2}
C(r) &=& r^2 \left (1 -  \frac{\cos (\sqrt{|\gamma|} r )}{ r\mathrm{ \sqrt{|\gamma|}}} \right )
\end{eqnarray}
for $\gamma <0$. {\color{black} The plots in Fig. \ref{plot-A1B1} and Fig. \ref{plot-A2B2} represent these metric potentials for $r\gtrsim 2$ (above the Schwarzschild horizon).} 

In Fig. \ref{plot-A1B1}, we depict the exact solutions of $A(r)$ (left panel) and $B(r)$ (right panel) for $n_B-n_F=-50, -500$ and $-2000$. This figure confirms 
the large-$r$ exponential behaviors of $A(r)$ in (\ref{metric-A1}) and $B(r)$ in (\ref{metric-B1}). It is clear that different $n_B-n_F$ curves rapidly merge and asymptote to unity -- the flat spacetime. In parallel with Fig. \ref{plot-A1B1}, we depict in Fig. \ref{plot-A2B2} the exact solutions of $A(r)$ (left panel) and $B(r)$ (right panel) for $n_B-n_F=50, 200$ and $2000$. Needless to say, this figure conforms with the large-$r$ sinusoidal behaviors of $A(r)$ in (\ref{metric-A2}) and $B(r)$ in (\ref{metric-B2}). It is clear that different $n_B-n_F$ values merge gradually (not rapidly) and asymptote to unity -- the flat spacetime. 

The exact numerical solutions above and their congruence with the asymptotic solutions show that the symmergent spacetime with the metric (\ref{varphi-eq}) asymptotes to flat spacetime in both $n_B-n_F < 0$ ($\gamma > 0$) and $n_B-n_F > 0$ ($\gamma < 0$) cases.  

\begin{figure}[h!]
   \centering
    \includegraphics[width=0.468\textwidth]{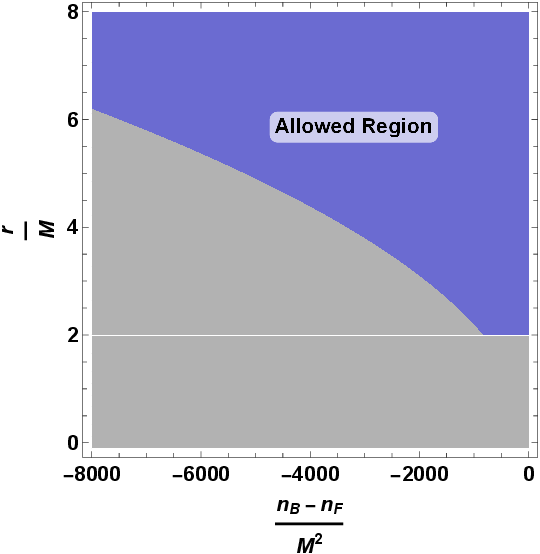} 
    \includegraphics[width=0.48\textwidth]{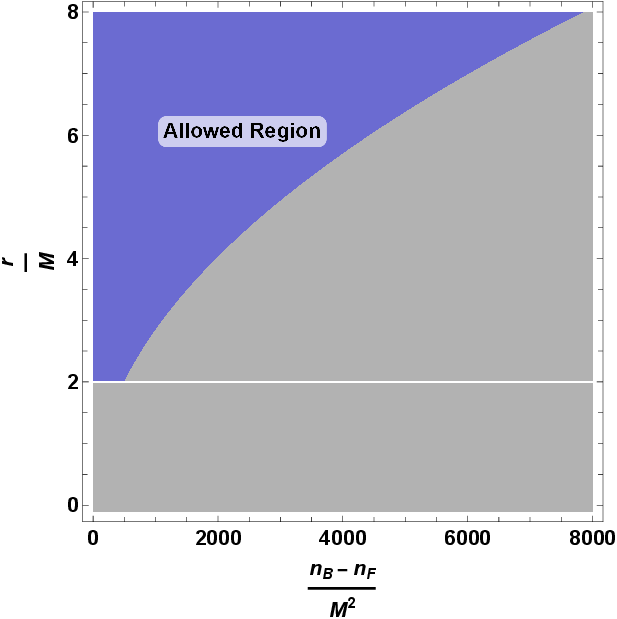}
    \caption{{\color{black} Allowed ranges of $r$ and $n_B-n_F$ after imposing $r\gtrsim 2$ from the validity of the metric (\ref{metric0}) and after imposing also $0 < A(r) < 1$ from causality. The left-panel (right-panel) corresponds to negative (positive) $n_B-n_F$ values. The white straight line at $r=2$ indicates the Schwarzschild horizon.}}
    \label{plot-allowedA}
\end{figure}

{\color{black}
In Fig. \ref{plot-allowedA}, we give the allowed regions in $r$ vs $n_B-n_F$ plane (shaded in blue). These regions are determined by imposing $r\gtrsim 2$ (see Fig. \ref{varphi_sols} above) from the validity of the metric (\ref{metric0}) and by imposing also $0 < A(r) < 1$ from the causality. The left-panel and right-panel correspond to $\gamma >0$  and $\gamma <0$, respectively. In the analyses below,  $r$ and $n_B-n_F$ values falling in the gray regions will not be considered. It is clear that the allowed region extends over a semi-infinite values in both $r$ and $n_B-n_F$ directions.}

\section{Tunneling of Symmerons and Other Particles: Hawking Temperature}

One of the key predictions of symmergence is that there necessarily exist new particles beyond the known SM spectrum, and these new particles do not have to interact non-gravitationally with the SM particles \cite{dark-sector1}. These new particles, the symmerons of Sec. \ref{subsec: new particles}, form a naturally-coupled sector \cite{dark-sector1}, and this sector could be visible (having SM-sized couplings to the SM), dark (having feeble interactions with the SM) or even black (having zero interactions with the SM). But, irrespective of their nature, symmerons couple to gravity and, to this end, black holes provide a viable environment to investigate their effects (on top of the $n_B-n_F$ dependence of $\gamma$ in (\ref{gamma})).

One way to observe symmerons is to measure the Hawking radiation from the symmergent black hole. More precisely, the evaporation rate of the symmergent black hole can reveal if there are light symmerons in the spectrum. Indeed, Hawking radiation is formed by the particles tunneling through the effective potential barrier at the horizon, and these particles can be the SM particles like the photon or featherweight symmerons. In fact, symmerons can form a dark Hawking radiation as they do not have to couple to the SM spectrum. 

Tunneling of particles from black holes and the resulting Hawking radiation have been studied for scalars \cite{Srinivasan:1998ty,Angheben:2005rm, Mitra:2006qa,Akhmedov:2006pg, Maluf:2018lyu,Gonzalez:2017zdz,Ovgun:2015jna,Ovgun:2017hje,Gecim:2013mfa,Ali:2007sh},  fermions \cite{Kerner:2007rr,Javed:2018msn,Rizwan:2018gpl,Chen:2013kha,Gecim:2019pft,Li:2008ws,Sharif:2012xq,Chen:2013tha} and vectors \cite{Kruglov:2014iya,Sakalli:2015taa,Javed:2019urt,Ovgun:2017iyb,Sakalli:2015jaa,Kuang:2017sqa,IbungochoubaSingh:2016prd}. 
Tunneling offers more than just a valuable method to verify the thermodynamic properties of black holes in that it also serves as an alternative conceptual approach to grasp the fundamental physical process behind black hole radiation. In general, tunneling methods involve the computation of the imaginary part of the action associated with the classically forbidden process of s-wave emission across the black hole horizon. This imaginary part is directly linked to the Boltzmann factor, which characterizes the probability of emission at the Hawking temperature. Here, we shall compute Hawking radiation as the tunneling of vector particles (like the photon and light vector symmerons) in static, spherically-symmetric, asymptotically-flat symmergent black holes with metric (\ref{metric-antz}). The motion equation  for an Abelian massive vector $V_\mu$ of mass $M_V$ is given by
\begin{eqnarray}
  \frac{1}{\sqrt{- g}} \partial_{\mu}  (\sqrt{- g} V^{\nu \mu}) -
  \frac{M_V^2 c^2}{\hbar^2} V^{\nu} = 0,
\label{Proca-eq}  
\end{eqnarray}
in which $V_{\nu \mu}=\partial_\nu V_\mu-\partial_\mu V_\nu$ is the field-strength tensor of the massive vector $V_\mu$. We will solve this equation by the WKB method \cite{WKB-0} and, to this end, we will suppress the speed of light $c$ but keep $\hbar$ to perform the WKB expansion. In fact, by labeling the components of $V_\mu$ as $V_\mu= (V_0, V_1, V_2, V_3)$, the Proca equation (\ref{Proca-eq}) can be decomposed as 
\begin{eqnarray}
\sqrt{\frac{B}{A}}\frac{1}{ C \sin \theta} \Bigg [ \partial_{r}\left(-\sqrt{\frac{B}{A}}  C \sin \theta \left(\partial_{t} V_{1}-\partial_{r} V_{0}\right)\right)+\partial_{\theta}\left(-\frac{\sin \theta}{\sqrt{A B}} \left(\partial_{t} V_{2}-\partial_{\theta} V_{0}\right)\right)  \left.  \right. \notag \\
+\partial_{\phi}\left(-\frac{1}{\sqrt{A B }\sin \theta} \left(\partial_{t} V_{3}-\partial_{\phi} V_{0}\right)\right) \Bigg]  - \frac{M_V^{2}}{\hbar^{2}}\left(-\frac{V_{0}}{A}\right)=0 , 
\label{eq1}
\end{eqnarray}

\begin{eqnarray}
 \sqrt{\frac{B}{A}} \frac{1}{C \sin \theta}  \Bigg[ \partial_t  \left(
\sqrt{\frac{B}{A}}  C \sin \theta  (\partial_t V_1 - \partial_r V_0) \right)
+ \partial_{\theta}  \left(\sqrt{A B} \sin \theta (\partial_r V_2
- \partial_{\theta} V_1) \right) 
  \notag \\ + \partial_{\phi}  \left( \frac{\sqrt{ AB}}{\sin \theta}  (\partial_r
V_3 - \partial_{\phi} V_1) \right) \Bigg ] - \frac{M_V^2}{\hbar^2} 
(B V_1) = 0 , \label{eq2}
\end{eqnarray}

\begin{eqnarray}
  \sqrt{\frac{B}{A}} \frac{1}{C \sin \theta}  \Bigg [ \partial_t  \left(
  \frac{\sin \theta}{\sqrt{A B}}  (\partial_t V_2 - \partial_{\theta} V_0)
  \right) + \partial_r  \left( - \sqrt{A B} \sin \theta (\partial_r
  V_2 - \partial_{\theta} V_1) \right)  
   \notag \\ + \partial_{\phi}  \left( \sqrt{\frac{A}{B}} \frac{1}{C \sin \theta}
  (\partial_{\theta} V_3 - \partial_{\phi} V_2) \right) \Bigg ] -
  \frac{m^2}{\hbar^2}  \left( \frac{V_2}{C}  \right) = 0 ,  
\label{eq3}
\end{eqnarray}
and
\begin{eqnarray}
\sqrt{\frac{B}{A}} \frac{1}{C \sin \theta} \Bigg [ \partial_t  \left(
\sqrt{\frac{1}{A B}} \frac{1}{\sin \theta}  (\partial_t V_3 - \partial_{\phi} V_0)
\right) + \partial_r  \left( - \frac{\sqrt{A B}}{\sin \theta} (\partial_r
V_3 - \partial_{\phi} V_1) \right)  
  \notag  \\ + \partial_{\theta}  \left( - \sqrt{\frac{A}{B}} \frac{1}{C \sin \theta} 
(\partial_{\theta} V_3 - \partial_{\phi} V_2) \right) \Bigg ] -
\frac{M_V^2}{\hbar^2}  \left( \frac{V_3}{C \sin^2 \theta} \right) = 0.  \label{eq4}
\end{eqnarray}
In the usual sense, the WKB wavefunction  \cite{WKB-0} 
\begin{eqnarray}
(V_0,V_1,V_2,V_3) = (v_0, v_1, v_2, v_3) \exp \left( \frac{i}{\hbar} I(t, r, \theta, \phi) \right),
\label{11}
\end{eqnarray}
can be determined order by order in $\hbar$ by expanding the action as 
\begin{eqnarray}
I (t, r, \theta, \phi) = I_0  (t, r, \theta, \phi) + \hbar I_1 (t, r, \theta, \phi) + \mathrm{O}\left(\hbar^2\right), 
\label{12}
\end{eqnarray}
in which $I_0$ is the classical action, $I_1$ is the one-loop correction, and so on. In the classical limit ($\hbar \rightarrow 0$), the classical action $I_0$ becomes the dominant piece (the semi-classical wavefunction) and admits the decomposition
\cite{Srinivasan:1998ty,Angheben:2005rm,Mitra:2006qa,Akhmedov:2006pg,Maluf:2018lyu}
\begin{eqnarray}
I = - \hbar \omega t + \hbar \int^r \kappa(r) dr,
\label{action-I}
\end{eqnarray}
as because it satisfies the Hamilton-Jacobi equation with energy $\hbar \omega=-\frac{\partial I_0}{\partial t}$ and momentum $\hbar \kappa(r)=\frac{\partial I_0}{\partial r}$. 


Under the decomposition (\ref{action-I}), the Proca equations (\ref{eq1}),(\ref{eq2}),(\ref{eq3}) and (\ref{eq4}) take the compact form
\begin{eqnarray}
(v_0, v_1, v_2, v_3)^T\cdot {\mathbb{C}} = 0,
\end{eqnarray}
in which the coefficient matrix
\begin{equation}
{\mathbb{C}} = 
\begin{pmatrix}
 \frac{\left(B \kappa^{2} + M_V^2 \right)}{A} & \frac{\omega  B \kappa} {A} & 0 & 0 \\
 \frac{ \omega  B \kappa}{A} & \frac{ B \left(\omega^2 - M_V^2 A \right)}{A} & 0 & 0 \\
 0 & 0 & \frac{\omega^2 - A \left(M_V^2 + B \kappa^{2}\right) }{A C} & 0 \\
 0 & 0 & 0 & \frac{\csc^2 \theta \left(\omega^2 - A \left(M_V^2 + B \kappa^{2}\right)\right)}{A C} 
\end{pmatrix}
\end{equation}
is set by the classical action in (\ref{action-I}). The vector field components ($v_0,v_1,v_2,v_3$) can take non-zero values if the coefficient matrix is not invertible, namely if it has zero determinant
\begin{align}
 {\rm Det}\left[{\mathbb{C}}\right]=0 \Longrightarrow \omega ^2 - A \left(M_V^2 + B \kappa^{2} \right) =0,
\end{align}
so that $\kappa^2=(\omega^2- M_V^2 A)/AB$. This solution for momentum $\hbar \kappa$ leads to the classical action 
\begin{align}
I_{0}(t,r) = \hbar \omega t \pm \hbar \int^r d{\tilde r} \sqrt{ \frac{ \omega^2 - M_V^2 A({\tilde r})}{A({\tilde r}) B({\tilde r})}},
\label{W-int}
\end{align}
whose value around the horizon $r=r_H$ should give the Euclidean action describing the barrier region. By definition, $A(r=r_H)=0$ and $B(r=r_H)=0$ at the horizon, and they can therefore be expanded thereabouts as
\begin{align}
A(r) &= A'(r_{H}) \left(r-r_{H}\right)+\mathrm{O}\left[\left(r-r_{H}\right)^{2}\right], \nonumber \\ 
B(r) &= B'(r_{H}) \left(r-r_{H}\right)+\mathrm{O}\left[\left(r-r_{H}\right)^{2}\right],
\end{align}
so that the classical action (\ref{W-int}) takes the form
\begin{equation}
I_0(t,r) =  \hbar \omega t \pm \hbar  \int  dr \frac{\omega}{\sqrt{A^{\prime}( r_H) B^{\prime}( r_H)} (r-r_{H})},
\label{W-int2}
\end{equation}
and its integration by the residue theorem gives
\begin{align}
I_0(t,r) = \hbar \omega t \pm \frac{ i \pi \hbar \omega}{\sqrt{A^\prime(r_H) B^\prime(r_H)}},
\end{align}   
with a $+$ ($-$) sign corresponding to outgoing (incoming) massive vector waves.  With this solution for the classical action, the WKB wavefunction in (\ref{11}) takes the form
\begin{eqnarray}
V_\mu = v_\mu \exp \left(i\omega t \mp
\frac{ \pi \omega}{\sqrt{A^\prime(r_H) B^\prime(r_H)}}\right),
\label{11p}
\end{eqnarray}
up to one-loop corrections, which we neglect. From this wavefunction follows the emission probability \cite{Kerner:2007rr}
\begin{eqnarray}
P_{em,\mu}=|v_\mu|^2 \exp\left\{-\frac{ 2\pi \omega}{\sqrt{A^\prime(r_H) B^\prime(r_H)}}\right\},
\end{eqnarray}
and the absorption probability
\begin{eqnarray}
P_{ab,\mu}=|v_\mu|^2 \exp\left\{\frac{ 2\pi \omega}{\sqrt{A^\prime(r_H) B^\prime(r_H)}}\right\},
\end{eqnarray}

\begin{figure}[h!]
   \centering
    \includegraphics[width=0.48\textwidth]{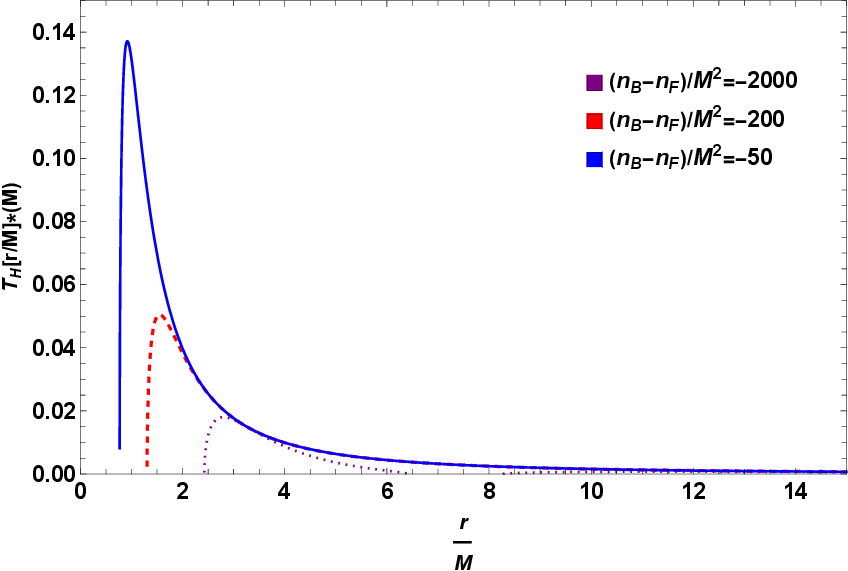} 
    \includegraphics[width=0.48\textwidth]{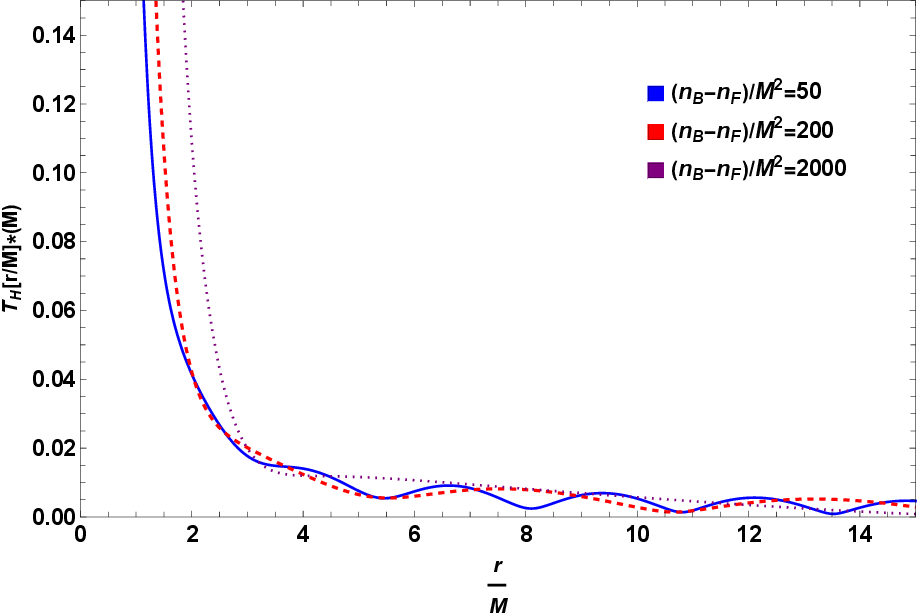}
    \caption{Hawking temperature as a function of the radial distance $r$ for selected negative values of $n_B - n_F$ ($\gamma >0$)(left panel) and positive values of $n_B - n_F$ ($\gamma < 0$)(right panel).}
    \label{plot-temperatures}
\end{figure}

One notes that the imaginary part of the action is the same for both incoming and outgoing solutions. Now, we want to define the decay rate of the black hole. To do that, we need to normalize the emission probability with respect to a given probability distribution. This is accomplished in reference \cite{Kerner:2006vu} by adding a constant part to the classical action and choosing that constant part in a way normalizing the absorption probability to unity \cite{Ovgun:2015jna,Ovgun:2016roz}. In our derivation, the same result is obtained by normalizing the emission probability to the absorption probability, namely
\begin{eqnarray}
\Gamma_\mu = \frac{P_{em,\mu}}{P_{ab, \mu}} = \exp\left\{-\frac{ 4\pi \omega}{\sqrt{A^\prime(r_H) B^\prime(r_H)}}\right\},
\label{prob-result}
\end{eqnarray}
and from this rate formula, one reads the Hawking temperature, which is inversely linked with the Boltzmann factor $\beta$,  \(T_{H}=\beta^{-1}\) as
\begin{align}
T_H =  \frac{\sqrt{A^\prime(r_H) B^\prime(r_H)}}{4 \pi },   
\end{align}
which differs from the Hawking temperature of Schwarzschild black hole by the presence of the conformal factor $\varphi(r)$ in the metric (\ref{metric0}). In fact, it takes the explicit form
\begin{align}
(T_H)_{\gamma > 0} = \frac{M}{ 2 \pi r_H^2} \sqrt{1 - \frac{e^{-2 \sqrt{\gamma} r_H}  \left(r_H + \sqrt{\gamma}  r_H^2 - 2 M ( 2 + \sqrt{\gamma}  r_H )  \right)^2}{4 M^2 \gamma r_H^2}},    
\end{align}
for $n_B<n_F$ ($\gamma > 0$), and  
\begin{align}
(T_H)_{\gamma \le 0} = \frac{M}{ 2 \pi r_H^2} \sqrt{1 - \frac{\left ( (r_H - 4 M) \cos ( \sqrt{|\gamma|} r_H ) + \sqrt{\gamma} r_H (r_H - 2 M) \sin ( \sqrt{|\gamma|} r_H )  \right )^2}{4 M^2 \gamma r_H^2}},   
\end{align}
for $n_B>n_F$ ($\gamma \leq 0$). We plot the Hawking temperature in Fig. \ref{plot-temperatures} as a function of radial distance $r$ for $n_B - n_F <0$ (left panel) and $n_B - n_F >0$ (right panel). As is seen from the left panel, when $n_B - n_F <0$, the Hawking temperature makes a peak and then falls off exponentially at large $r$ such that the larger the $|n_B - n_F|$ smaller the peak temperature and farther the peak position. These peaks can indicate negative $n_B - n_F$ if detected by black hole observations. 

The Hawking temperature for $n_B - n_F >0$, as indicated by the right panel of  Fig. \ref{plot-temperatures}, decreases with $r$ exhibiting a periodic pattern. The fall is gradual. Thus, it becomes easier to distinguish different $n_B - n_F$ values compared to $n_B - n_F <0$ case in the left panel. The periodic behavior can serve as the indicator of positive $n_B - n_F$ if it can be detected by black hole observations. 

\begin{figure}[h!]
   \centering
    \includegraphics[width=0.467\textwidth]{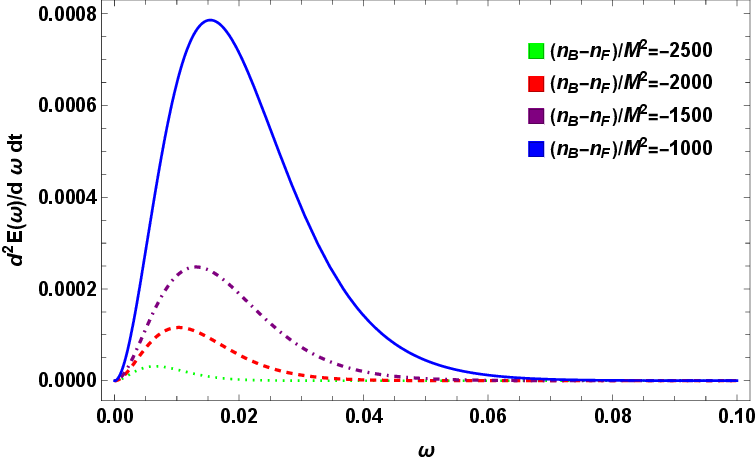} 
    \includegraphics[width=0.460\textwidth]{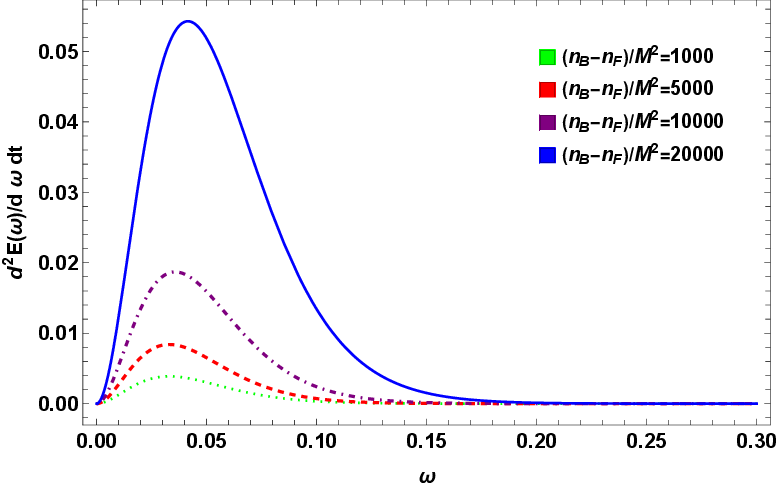}
    \caption{The energy emission rate in \eqref{energy-emission} as a function of frequency for $n_B - n_F<0$ (left panel) and $n_B - n_F>0$ (right panel). The movement of the peak position is a distinctive feature.}
    \label{fig-emission-rate-1}
\end{figure}

The vector field Hawking temperature in Fig. \ref{plot-temperatures} illustrates the contribution of one single flavor.  But, in general, black holes emit not only the photons but also neutrinos \cite{Page1976}. In symmergent gravity, the emission spectrum can be multifarious as it can now involve symmerons \cite{demir0,demir1,demir2}. Indeed, as discussed in detail in Sec. \ref{subsec: new particles}, symmerons can be dark (or even black), and it can be hard to detect them at collider experiments or dark matter sector searches. But, irrespective of their nature, they can contribute to the Hawking radiation and affect the evaporation rate of the black hole. To be definitive, the Hawking emission rate of any particle species $i$ is given by \cite{Page1976,Baker2022,Cheek2021,Hawking1975,Decanini2011}
\begin{align}
\frac{d^2 N_i}{d \omega dt} = \frac{n_i \sigma_i \omega^2}{2\pi^2 \left ( e^{\omega/T_H}  \pm 1 \right )}, 
\label{emission-rate}
\end{align}
in which $n_i$ is the number of degrees of freedom of the species $i$, $+(-)$ stands for fermions (bosons), $\sigma_i$ is the absorption cross section, and $T_H$ is the Hawking temperature of the black hole. This rate conforms with the black hole decay rate in  (\ref{prob-result}) except for the quantum-statistics factors. The absorption cross section $\sigma_i$ is given by $\sigma_i = \pi E^{-2} \Gamma_i$, where $\Gamma_i$ is the greybody factor \cite{Liao2013}. The greybody factor is a measure of deviations from the perfect black body spectrum in (\ref{prob-result}) and results from the backscattering of the particles under the strong gravity of the black hole. To a good approximation, the absorption cross-section $\sigma_i$ can be identified with the geometrical cross-section of the black hole's photon sphere \cite{Mashhoon1973}
\begin{align}
\sigma_{i} \approx \pi R_{sh}^2  
\label{xsection}
\end{align}
as a universal formula for all species. In parallel with the change of the particle number in (\ref{emission-rate}), a black hole emits energy $E$ at a rate \cite{Wei2013}
\begin{align}
\frac{d^2 E}{d \omega dt} = \frac{2 \pi^2\sigma_{i} \omega^3}{\left ( e^{\omega/T_H}  \pm 1 \right ) } 
\label{energy-emission}
\end{align}
in which the absorption cross section $\sigma_i$ can be identified with  (\ref{xsection}). This energy emission rate we plot in Fig. \ref{fig-emission-rate-1} for $n_B-n_F<0$ (left panel) and $n_B-n_F>0$ (right panel). It is clear that the energy emission rate makes a peak a specific frequency such that when $n_B-n_F$ decreases in a negative direction, the peak gets suppressed and shifts to smaller $\omega$ values. In contrast, when $n_B-n_F$ increases in a positive direction, the peak gets pronounced and shifts to larger $\omega$ values. 
 
\begin{figure}[h!]
   \centering
    \includegraphics[width=0.467\textwidth]{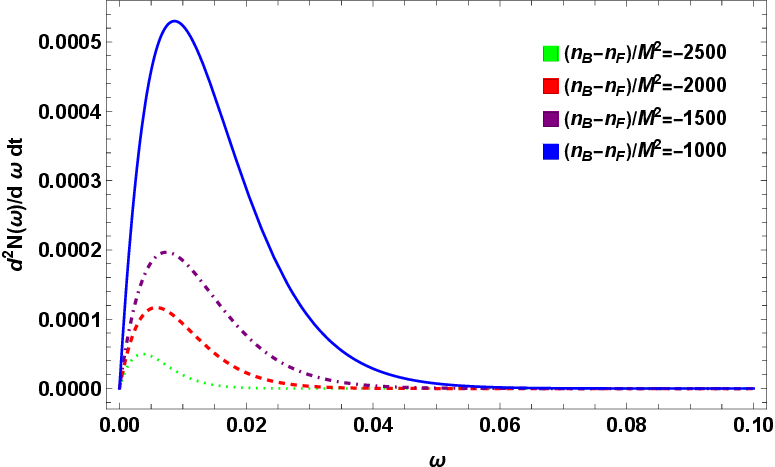} 
    \includegraphics[width=0.463\textwidth]{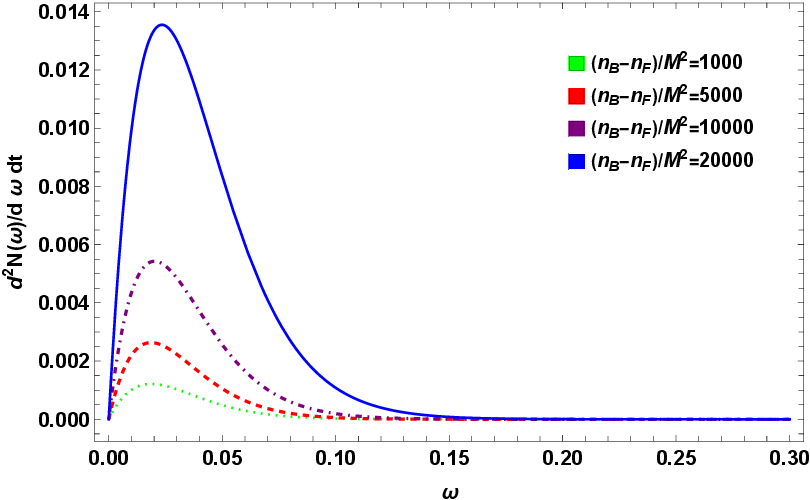}
    \caption{The particle emission rate in \eqref{emission-rate} as a function of frequency for $n_B - n_F<0$ (left panel) and $n_B - n_F>0$ (right panel). The movement of the peak position is a distinctive feature.}
    \label{fig-emission-rate-2}
\end{figure}

Depicted in Fig. \ref{fig-emission-rate-2} is the particle emission rate as a function of the frequency $\omega$ for $n_B-n_F<0$ (left panel) and $n_B-n_F>0$ (right panel). It exhibits a similar behavior as the energy emission rate in Fig. \ref{fig-emission-rate-1} in that the rate at which a given particle species is emitted makes a peak at a specific frequency. The peak decreases with the decreasing values of negative $n_B-n_F$ and shifts to lower frequencies. In parallel, the peak increases with the increasing values of positive $n_B-n_F$ and shifts to higher frequencies.  It is clear that the positions of the peaks could give information about whether or not the bosonic degrees of freedom in the Universe are larger than the fermionic ones. It is worth noting that positive $n_B-n_F$ should be significantly large for the peaks to show noticeable shifts (the right panels of both 
Fig. \ref{fig-emission-rate-1} and Fig. \ref{fig-emission-rate-2}).

In view of the symmergent gravity explored, it proves useful to explicate the symmeron contributions, and, to this end, the total particle emission rate proves to be a useful quantity. Indeed, as follows from the individual rates in (\ref{emission-rate}), the total emission rate consists of all species of any spin \cite{Page1976}
\begin{align}
\frac{dN}{dt} = \sum_{i=\gamma,\nu, symmerons} \int \frac{n_i \sigma_i \omega^2 d\omega}{\pi^2 \left ( e^{\omega/T_H}  \pm 1 \right )} \approx \sum_{i=\gamma,\nu, symmerons} \!\!\!\frac{k_i \zeta(3)}{\pi} n_i T_H^3 R_{sh}^2 = \frac{\zeta(3)}{\pi} T_H^3 R_{sh}^2 \left(13 + 2 n_{B}^{(0)} + \frac{3}{2} n_{F}^{(0)} \right)
\label{emission-rate-tot}
\end{align}
with the dominance of light (massless) particles as because contributions of massive particles get suppressed as $e^{-mass/T_H}$. In this expression, $\zeta(3)\approx 1.202$ (Riemann zeta-function) and $k_i=2 (3/2)$ for bosons (fermions). In the last equality, photon ($n_i=2$) plus neutrino ($n_i=6$) contributions sum up to $13$. This SM rate is extended by the contributions of  $n_{B}^{(0)}$ massless bosonic  symmerons plus $n_{F}^{(0)}$ massless  fermionic symmerons. It is all clear that symmerons increase the particle emission rate (black hole evaporation) in direct proportion to number of massless symmerons. This effect on the total emission rate might be determined or inferred by observing a given black hole at its different phases of evolution in a certain environment. One thus concludes that possible observation of the black hole evaporation can give information about new massless/light particles beyond the known particles in the SM. And symmergent gravity predicts the existence of such particles and also predicts that these particles may not interact at all with the known particles non-gravitationally \cite{demir0,demir1,demir2}. 

\section{Photon Sphere, Shadow Cast and Weak Deflection Angle of the Symmergent Black Hole} \label{sec2}

\begin{figure}[h!]
   \centering
    \includegraphics[width=0.465\textwidth]{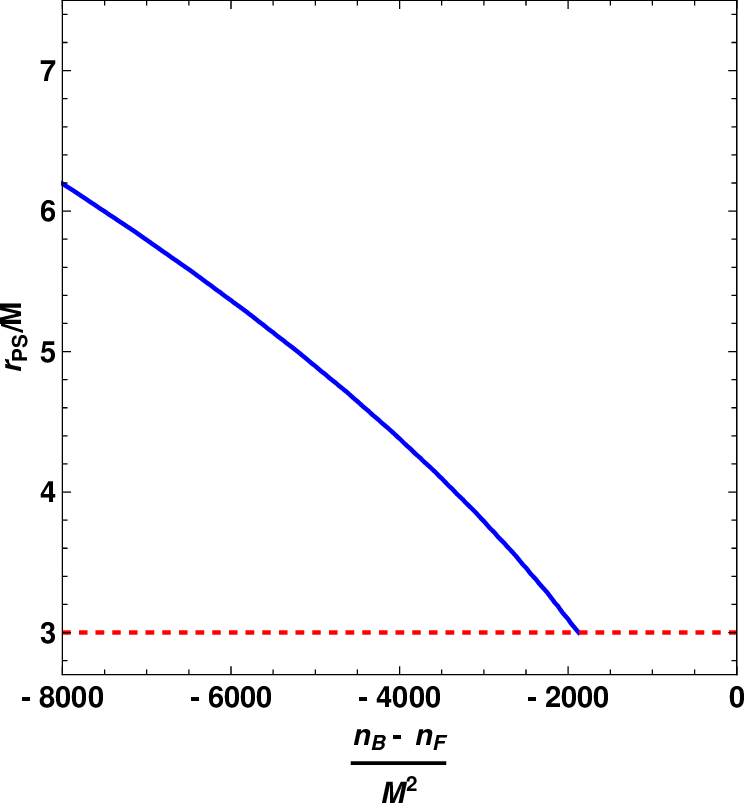} 
    \includegraphics[width=0.48\textwidth]{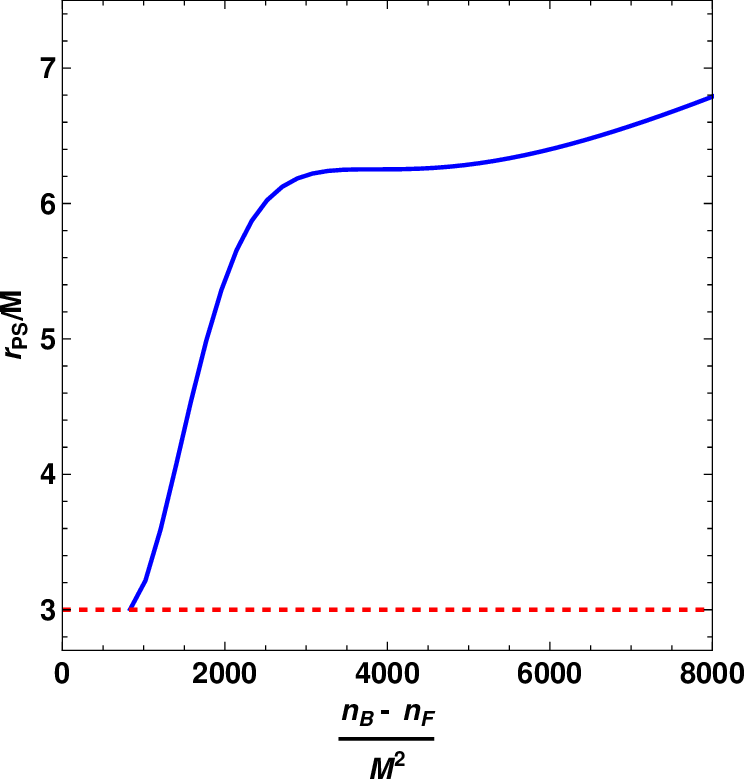}
    \caption{{\color{black}The photon sphere behavior as a function of  $n_B - n_F$ for $n_B - n_F<0$ (left panel) and  $n_B - n_F>0$ (right panel). The red-dashed line at $r=3M$ is the photon sphere radius for the Schwarzschild metric. The blue line, falling in the allowed domains in Fig. \ref{plot-allowedA}, are the photon sphere radius characterizing the non-Schwarzschild part of the metric (\ref{metric0}).}}
    \label{photon-radius}
\end{figure}

In this section, we analyze the photon sphere and shadow cast of the symmergent black hole. Given the spherical symmetry of the metric in (\ref{metric0}), it suffices to analyze null geodesics in the equatorial plane ($\theta=\pi/2$).  The null geodesic Lagrangian
\begin{equation}
    \mathcal{L} = \frac{1}{2} g_{\mu\nu} \dot{x}^\mu \dot{x}^\nu = \frac{1}{2}\left( -A(r) \dot{t}^2 +\frac{\dot{r}^2}{B(r)}  + C(r) \dot{\phi}^2 \right),
\end{equation}
leads to the energy 
\begin{equation}
    E = A(r)\frac{dt}{d\lambda},
    \label{energy}
\end{equation}
and the angular momentum 
\begin{equation}
   L = C(r)\frac{d\phi}{d\lambda},
   \label{angular-mom}
\end{equation}
as two constants of motion. Their ratio gives the impact parameter 
\begin{equation}
    b =\frac{L}{E} = \frac{C(r)}{A(r)}\frac{d\phi}{dt},
\end{equation}
involving the coordinate angular velocity $d\phi/dt$. Using the impact parameter $b$, the null geodesic $ds^2=0$ enables us to relate $r$ to $\phi$ as \cite{Perlick:2021aok}
\begin{align}
    \left(\frac{dr}{d\phi}\right)^2 =B(r) C(r)\left(\frac{h(r)^2}{b^2}-1\right),
    \label{r-phi}
\end{align}
after defining
\begin{equation} \label{eh(r)}
    h(r)^2 = \frac{C(r)}{A(r)},
\end{equation}
for simplicity. The null geodesic $r=r(\phi)$ remains stable if these two conditions are satisfied: The first condition is $dr/d\phi=0$, and it implies
$h(r_{\rm PS})=b$ as follows from (\ref{r-phi}). The second condition is $d^2r/d\phi^2=0$, and it requires  $\frac{d}{d r}\left(\frac{A(r)}{C(r)}\right)_{|r=r_{\rm PS}}=0$ as follows again from (\ref{r-phi}). These two conditions lead to the relation 
\begin{eqnarray}
C(r_{\rm PS})A'(r_{\rm PS})-C'(r_{\rm PS})A(r_{\rm PS})=0,
\label{eq:photonspherecomplete}
\end{eqnarray}
as the equation defining the radius of stable sphere $r=r_{\rm PS}$. {\color{black}  For both $\gamma > 0$ and $\gamma < 0$, one solution of equation \eqref{eq:photonspherecomplete} is $r_{\rm PS}=3M$, which corresponds to the Schwarzschild part of the metric (\ref{metric0}). The other solution is a nontrivial function of $\gamma$, and it reads  $r_{\rm PS} = \text{LambertW}(1)\sqrt{\gamma}$ for $\gamma>0$. For $\gamma<0$, a closed analytic solution is not available and it becomes necessary to resort to numerical techniques.}  
The two cases are depicted in Fig. \ref{photon-radius} for both $n_B - n_F<0$ (left panel) and $n_B - n_F>0$ (right panel). As follows from the left panel, the second photon sphere occurs at larger and larger radii for smaller and smaller negative $n_B-n_F$. The right panel, on the other hand, reveals that the second photon sphere occurs at a radius $r_{\rm PS}\approx 6.3 M$ at large $n_B-n_F$. Consequently, photons form two stable spheres for both $n_B - n_F<0$ and $n_B - n_F>0$, and the radii of these spheres can be a distinctive signature of the symmergent gravity.

\begin{figure}[h!]
   \centering
    \includegraphics[width=0.45\textwidth]{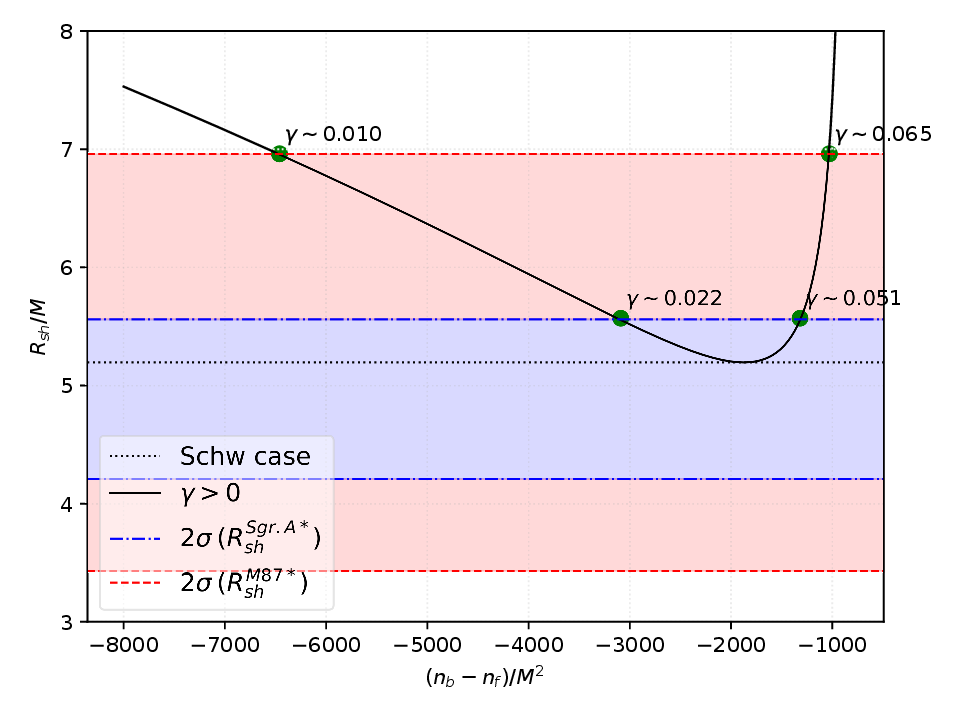}
    \includegraphics[width=0.45\textwidth]{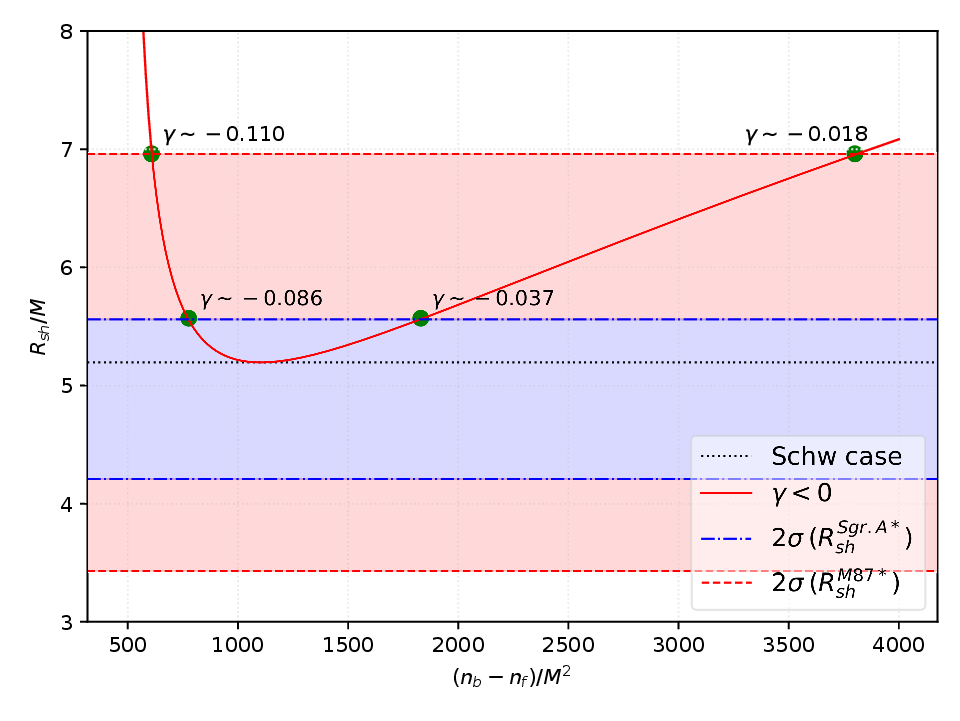}
    \caption{{\color{black}The radius of the shadow of the symmergent black hole for $n_B - n_F<0$ (left panel) and $n_B - n_F>0$ (right panel). The dashed line is the shadow of the Schwarzschild black hole. Superimposed $1\sigma$ and $2\sigma$ bands follow from the Sgr. A* observational data in Table  \ref{tab_obs}. The minimal shadow radii  ($R_{\rm sh}=3\sqrt{3}$) around $n_b-n_f\approx -2000$ (left) and $n_b-n_f\approx 1100$ (right) fall into the allowed domains (blue) in Fig. \ref{plot-allowedA}.}}
    \label{shadow-radius}
\end{figure}

The non-rotating symmergent black hole under concern can produce only a circular shadow silhouette. The shadow is caused by photons falling within the photon sphere of radius $r=r_{\rm PS}$. For the Schwarzschild black hole, the photon radius is $r_{\rm PS}=3 M$, and its shadow has the radius  $R_{\rm sh} = 3\sqrt{3}M$. The shadow radius coincides with the critical impact parameter $b=b_{\rm crit}$ at which photons are just at the threshold of getting caught by the gravity of the black hole. Needless to say, $R_{\rm sh}$ is nothing but the gravitationally lensed image of the photon sphere of radius $r=r_{\rm PS}$ \cite{EventHorizonTelescope:2020qrl}. {\color{black} In symmergent gravity, the shadow radius is given by 
\begin{eqnarray}
R_{\rm sh} = \sqrt{\frac{C(r_{\rm PS})}{A(r_{\rm PS})}}
\label{eq:rshcomplete}
\end{eqnarray}
and it can be determined analytically for $\gamma>0$ and $\gamma <0$ separately. However, the analytic expressions are too lengthy to be suggestive and therefore it proves convenient to proceed with the numerical calculations. 
It is clear that $R_{\rm sh}$ varies with $n_B-n_F$ through the photon sphere radius $r_{\rm PS}$ in Fig. \ref{photon-radius}.} We plot in Fig. \ref{shadow-radius} the shadow radius of the symmergent black hole as a function of $n_B-n_F$ for $n_B-n_F<0$ (left panel) and $n_B-n_F>0$ (right panel). The two panels are nearly mirror-symmetric to each other. (Actually, the $R_{\rm sh}$ lies in different $n_B-n_F$ ranges.) Superimposed on the $R_{\rm sh}$ plot in Fig. \ref{shadow-radius} are the observational results on the Sgr. A* tabulated in Table \ref{tab_obs}. (The bounds from M87* are much milder \cite{Vagnozzi:2022moj}.) The observation of the exponential and inverse relationship displayed by $n_B-n_F$ continues to captivate our interest, regardless of the variation range of the parameters. In Fig. \ref{shadow-radius}, we present the upper boundaries of $|n_B-n_F|$; these values have been contextualized based on the Event Horizon Telescope (EHT) observations. As per the statistical interpretation at a confidence level (C.L.) of 
\(68\%\) as delineated in the study by Vagnozzi et al. (2022) \cite{Vagnozzi:2022moj}, the upper constraint for $n_B-n_F$ manifests as $-1700$ and $-2100$ for the left diagram and $1300$ and $1000$ for the right diagram. As the figure reveals, symmergent effects fall in the $2\sigma$ band, with a tiny tail in the $1\sigma$ band. {\color{black} The $R_{\rm sh}$ nears the Schwarzchild value of $R_{\rm sh}=3\sqrt{3}$ around $n_b-n_f\approx -2000$ (left) and $n_b-n_f\approx 1100$ (right). These minimal values fall into the allowed domains (blue) in Fig. \ref{plot-allowedA}. In view of these small $1\sigma$ tails, future higher-precision observations on the supermassive and other black holes could be expected to bound the symmergent parameters.} 

\begin{table}
    \centering
    \begin{tabular}{ |p{2cm}| p{3cm}| p{4cm}| p{2cm}|p{3.8cm}|}
    \hline
    Black hole & Mass ($M_\odot$) & Angular diameter: $2\alpha_\text{sh}$ ($\mu$as) & Distance (kpc) & Schw. dev. ($2\sigma$) \\
    \hline
    Sgr. A*   & $4.3 \pm 0.013$x$10^6$ (VLTI)    & $48.7 \pm 7$ (EHT) &   $8.277 \pm 0.033$ & $3\sqrt{3}[1 + (-0.06 \pm 0.065) ]$\\
    \hline
    M87* &   $6.5 \pm 0.90$x$10^9$  & $42 \pm 3$   & $16800$ & $3\sqrt{3}(1 \pm 0.17 )$ \\
    \hline
    \end{tabular}
    \caption{Observational constraints on the mass, angular shadow radius, distance, and Schwarzschild deviation of the supermassive black holes Sgr. A* and M87*.}
    \label{tab_obs}
\end{table}

\begin{figure}[htp!]
   \centering
    \includegraphics[width=0.45\textwidth]{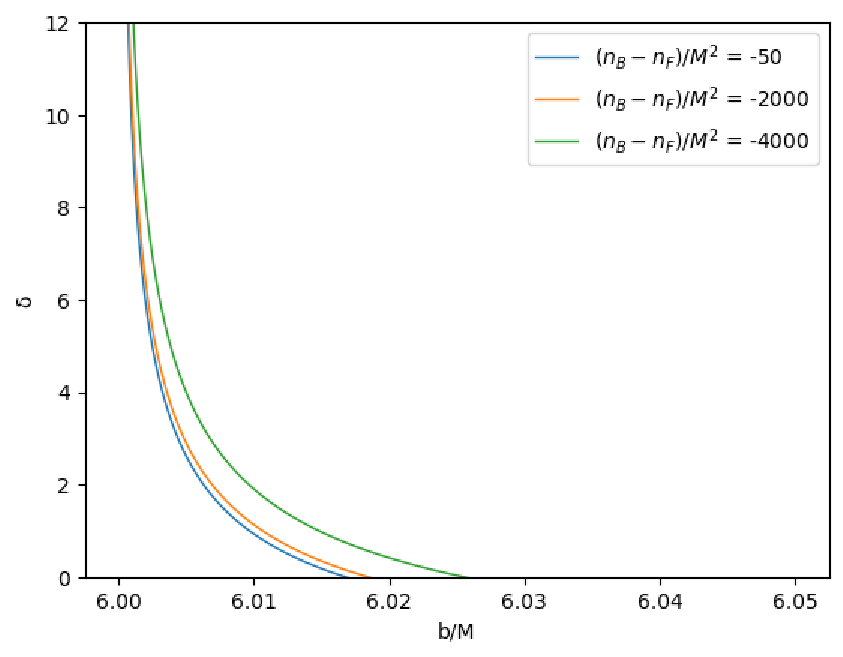}
    \includegraphics[width=0.453\textwidth]{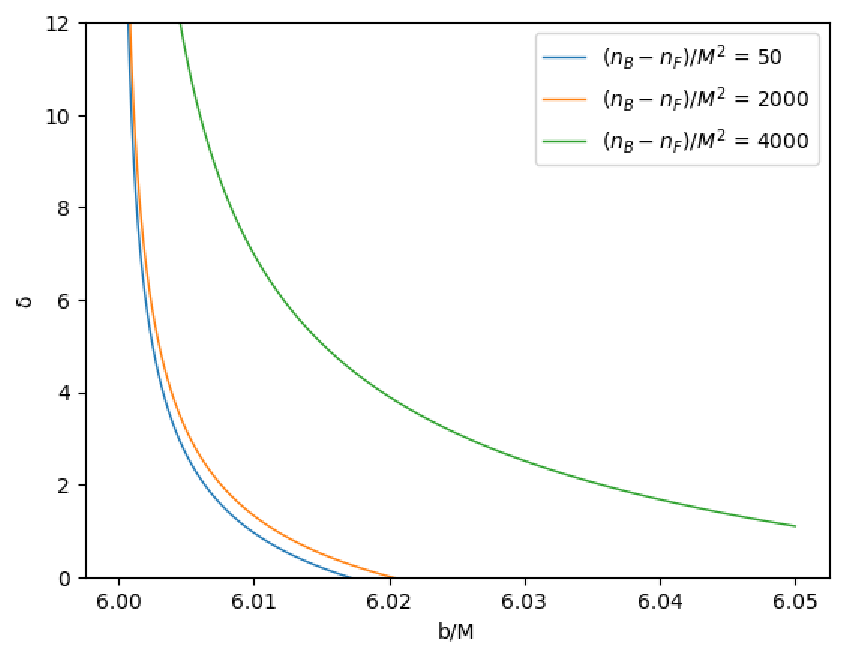}
    \caption{Weak deflection angle from the symmergent black hole as a function of the impact parameter for $n_B - n_F<0$ (left panel) and $n_B - n_F>0$ (right panel). }
    \label{WDA}
\end{figure}

Lastly, we analyze the deflection angle numerically in weak field limits from the symmergent black hole described by the metric (\ref{metric0}). For the symmergent metric (\ref{metric0}), the bending angle in gravitational lensing is given by \cite{Weinberg:1972kfs,Lu:2019ush,Virbhadra:1998dy} 
\begin{equation}
\delta \left(r_{0}\right)= \int_{r_{0}}^{\infty} \frac{2}{\sqrt{B(r) C(r)} } \frac{\mathrm{~d} r}{\sqrt{\frac{C(r)}{C\left(r_{0}\right)} \frac{A\left(r_{0}\right)}{A(r)}-1}} -\pi,
\end{equation}
where $r_0$ is the closest approach distance of the light ray to the symmergent black hole. In the weak deflection lensing, $r_0\gg 2M$ so that $\delta \ll 1$. We perform this integration numerically and plot the results in Fig. \ref{WDA} for both $n_B-n_F<0$ (left panel) and $n_B-n_F>0$ (right panel)
As the figure suggests, the deflection angle is similar in the two cases, with a rapid fall-off with the impact parameter. It is with fairly small impact parameters ($b\gtrsim 6$) that one starts discriminating different $n_B-n_F$ values. 

\section{Conclusion} \label{sec7}

Symmergent gravity is an emergent gravity theory with $R+R^2$ curvature sector and a new particle sector. In the present work, we have constructed and analyzed asymptotically flat, static, spherically-symmetric symmergent gravity black holes. In the case of constant scalar curvature, the quadratic curvature term (coefficient of $R^2$) does not affect the asymptotically flat spacetimes \cite{constant-R-1,constant-R-2,constant-R-3}. In the case of variable scalar curvature, however, there arise asymptotically-flat solutions with explicit dependence on the quadratic curvature term  \cite{nguyen3} (see also \cite{nguyen1,nguyen2}). In the present work, we have studied such variable-scalar-curvature black holes in detail. 

In Sec. II, we have a detailed discussion of the symmergent gravity \cite{demir0,demir1} regarding its new particles (symmeron) sector (Sec. IIA) and its curvature sector (Sec. IIB). In Sec. III, we have explicitly constructed asymptotically-flat, static, spherically-symmetric symmergent gravity black holes with variable scalar curvature. Our analysis goes beyond that in \cite{nguyen3} as we considered both positive and negative values of the boson-fermion number difference. In both cases, we have shown the asymptotic flatness of the metric, with approximate analytic calculations and the exact numerical solutions. In Sec. IV, we have computed the Hawking temperature using the tunneling method and concluded that black hole evaporation could be accelerated if there exist light symmerons of a significant number. In Sec. V, we have analyzed how the symmergent black hole can be probed via its shadow cast and weak deflection angle. We have shown that symmergent effects on the shadow are essentially a $2-\sigma$ effect, and the weak deflection angle can distinguish different boson-fermion number differences at fairly large impact factors. 

Our analysis in this work of the symmergent gravity is completely new in view of its asymptotic flatness and in view also of its sensitivity to the quadratic curvature term (coefficient of $R^2$). The analysis here can be extended to other black hole properties like quasinormal modes and grey body factors. The analysis here can also be extended by iterating Nguyen's solution to quadratic and higher-powers of the conformal factor.

\acknowledgments 
The work of B. P. is supported by Sabanc{\i} University, Faculty of Engineering and Natural Sciences by Faculty Postdoctoral Researcher Grant. B. P., R. P. and A. {\"O}.  would like to acknowledge networking support by the COST Action CA18108 - Quantum gravity phenomenology in the multi-messenger approach (QG-MM). B. P., A. {\"O}. and D. D.  would like to acknowledge networking support by the COST Action CA21106 - COSMIC WISPers in the Dark Universe: Theory, astrophysics and experiments (CosmicWISPers).

\bibliography{references}
\end{document}